\documentclass[twocolumn,times]{aastex63}
\shorttitle{hst observations of the muse ultra deep field}
\shortauthors{Revalski et al.}
\graphicspath{{./}{figures/}}

\usepackage{float}
\usepackage{color}
\usepackage{comment}
\usepackage{graphics}
\usepackage{graphicx}
\usepackage{epstopdf}
\usepackage{microtype}
\usepackage{subfigure}
\usepackage[normalem]{ulem}
\usepackage{natbib}
\usepackage{amsmath}
\definecolor{malachite}{rgb}{0.04, 0.85, 0.32}

\usepackage{soul,xcolor}
\setstcolor{red}

\NewPageAfterKeywords

\begin{document}

\title{The MUSE Ultra Deep Field (MUDF). III. Hubble Space Telescope WFC3 Grism Spectroscopy and Imaging}


\correspondingauthor{Mitchell Revalski}
\email{mrevalski@stsci.edu}

\author[0000-0002-4917-7873]{Mitchell Revalski}
\affiliation{Space Telescope Science Institute, 3700 San Martin Drive, Baltimore, MD 21218, USA}

\author[0000-0002-9946-4731]{Marc Rafelski}
\affiliation{Space Telescope Science Institute, 3700 San Martin Drive, Baltimore, MD 21218, USA}
\affiliation{Department of Physics and Astronomy, Johns Hopkins University, Baltimore, MD 21218, USA}

\author[0000-0001-6676-3842]{Michele Fumagalli}
\affiliation{Dipartimento di Fisica G. Occhialini, Universit\`a degli Studi di Milano-Bicocca, Piazza della Scienza 3, I-20126 Milano, Italy}
\affiliation{INAF - Osservatorio Astronomico di Trieste, via G. B. Tiepolo 11, I-34143 Trieste, Italy}

\author[0000-0002-9043-8764]{Matteo Fossati}
\affiliation{Dipartimento di Fisica G. Occhialini, Universit\`a degli Studi di Milano-Bicocca, Piazza della Scienza 3, I-20126 Milano, Italy}
\affiliation{INAF - Osservatorio Astronomico di Brera, via Bianchi 46, I-23087 Merate (LC), Italy}

\author[0000-0003-3382-5941]{Norbert Pirzkal}
\affiliation{Space Telescope Science Institute, 3700 San Martin Drive, Baltimore, MD 21218, USA}

\author[0000-0003-3759-8707]{Ben Sunnquist}
\affiliation{Space Telescope Science Institute, 3700 San Martin Drive, Baltimore, MD 21218, USA}

\author[0000-0002-0604-654X]{Laura J. Prichard}
\affiliation{Space Telescope Science Institute, 3700 San Martin Drive, Baltimore, MD 21218, USA}

\author[0000-0002-6586-4446]{Alaina Henry}
\affiliation{Space Telescope Science Institute, 3700 San Martin Drive, Baltimore, MD 21218, USA}

\author[0000-0002-9921-9218]{Micaela Bagley}
\affiliation{Department of Astronomy, The University of Texas at Austin, Austin, TX 78712, USA}

\author[0000-0002-6095-7627]{Rajeshwari Dutta}
\affiliation{Dipartimento di Fisica G. Occhialini, Universit\`a degli Studi di Milano-Bicocca, Piazza della Scienza 3, I-20126 Milano, Italy}
\affiliation{INAF - Osservatorio Astronomico di Brera, via Bianchi 46, I-23087 Merate (LC), Italy}

\author[0000-0002-5810-318X]{Giulia Papini}
\affiliation{Dipartimento di Fisica G. Occhialini, Universit\`a degli Studi di Milano-Bicocca, Piazza della Scienza 3, I-20126 Milano, Italy}

\author[0000-0002-4770-6137]{Fabrizio Arrigoni Battaia}
\affiliation{Max-Planck-Institut f\"ur Astrophysik,
Karl-Schwarzschild-Strasse 1, D-85748 Garching bei M\"unchen, Germany}

\author[0000-0003-3693-3091]{Valentina D'Odorico}
\affiliation{INAF - Osservatorio Astronomico di Trieste, via G. B. Tiepolo 11, I-34143 Trieste, Italy}
\affiliation{Scuola Normale Superiore, Piazza dei Cavalieri 7, I-56126, Pisa, Italy}
\affiliation{IFPU - Institute for Fundamental Physics of the Universe, via Beirut 2, I-34151 Trieste, Italy}

\author[0000-0001-8460-1564]{Pratika Dayal}
\affiliation{Kapteyn Astronomical Institute, University of Groningen, P.O. Box 800, 9700 AV Groningen, The Netherlands}

\author[0000-0001-8489-2349]{Vicente Estrada-Carpenter}
\affiliation{Department of Astronomy \& Physics, Saint Mary's University, 923 Robie Street, Halifax, NS, B3H 3C3, Canada}

\author[0000-0002-1209-9680]{Emma K. Lofthouse}
\affiliation{Dipartimento di Fisica G. Occhialini, Universit\`a degli Studi di Milano-Bicocca, Piazza della Scienza 3, I-20126 Milano, Italy}

\author[0000-0003-0083-1157]{Elisabeta Lusso}
\affiliation{Dipartimento di Fisica e Astronomia, Universit\`a di Firenze, via G. Sansone 1, I-50019 Sesto Fiorentino, Firenze, Italy}
\affiliation{INAF -- Osservatorio Astrofisico di Arcetri, Largo Enrico Fermi 5, I-50125 Firenze, Italy}

\author[0000-0003-4866-110X]{Simon L. Morris}
\affiliation{Centre for Extragalactic Astronomy, Durham University, South Road, Durham DH1 3LE, UK}

\author[0000-0001-5294-8002]{Kalina~V.~Nedkova} 
\affiliation{Department of Physics and Astronomy, Johns Hopkins University, Baltimore, MD 21218, USA}

\author[0000-0001-7503-8482]{Casey Papovich}
\affiliation{Department of Physics and Astronomy, Texas A\&M University, College Station, TX 77843-4242, USA}
\affiliation{George P.\ and Cynthia Woods Mitchell Institute for Fundamental Physics and Astronomy, Texas A\&M University, College Station, TX 77843-4242, USA}

\author[0000-0002-4288-599X]{Celine Peroux}
\affiliation{European Southern Observatory, Karl-Schwarzschild-Strasse 2, D-85748 Garching bei M{\"u}nchen, Germany}
\affiliation{Aix Marseille Universit\'e, CNRS, LAM (Laboratoire d'Astrophysique de Marseille) UMR 7326, F-13388, Marseille, France}

\begin{abstract}
We present extremely deep Hubble Space Telescope (HST) Wide Field Camera 3 (WFC3) observations of the MUSE Ultra Deep Field (MUDF). This unique region of the sky contains two quasars at $z \approx$~3.22 that are separated by only $\sim$500~kpc, providing a stereoscopic view of gas and galaxies in emission and absorption across $\sim$10 billion years of cosmic time. We have obtained 90~orbits of HST WFC3 G141 near-infrared grism spectroscopy of this field in a single pointing, as well as 142 hours of optical spectroscopy with the Very Large Telescope (VLT) Multi Unit Spectroscopic Explorer (MUSE). The WFC3 (F140W, F125W, and F336W) and archival WFPC2 (F702W and F450W) imaging provides five-filter photometry that we use to detect 3,375 sources between $z \approx$~0~--~6, including 1,536 objects in a deep central pointing with both spectroscopic and photometric coverage. The F140W and F336W mosaics reach exceptional depths of $m_\mathrm{AB}\approx$~28 and 29, respectively, providing near-infrared and rest-frame ultraviolet information for 1,580 sources, and we reach 5$\sigma$ continuum detections for objects as faint as $m_\mathrm{AB}\approx$~27 in the grism spectra. The extensive wavelength coverage of MUSE and WFC3 allows us to measure spectroscopic redshifts for 419 sources, down to galaxy stellar masses of log(M/M$_{\odot}$)~$\approx$~7 at $z \approx$~1~--~2. In this publication, we provide the calibrated HST data and source catalogs as High Level Science Products for use by the community, which includes photometry, morphology, and redshift measurements that enable a variety of studies aimed at advancing our models of galaxy formation and evolution in different environments.
\end{abstract}



\section{Introduction}\label{sec:intro}

\subsection{Scientific Motivation}\label{ssec:motivation}

Understanding the connections between galaxies and their surrounding gaseous environments is fundamental to advancing our theoretical models of galaxy formation and evolution \citep{Conselice2014, Somerville2015, Rahmati2016, Naab2017, Peeples2019, Corlies2020, Oppenheimer2020, Lochhaas2021}. Studying these relationships, and how they change over cosmic time, requires sensitive multiwavelength observations of galaxies and their surrounding gas over a large range in redshift \citep{Chen2010, Rubin2018, Dutta2021}. While high-mass galaxies produce the majority of the light that we observe, low-mass galaxies at the faint end of the luminosity function dominate the population. These numerous low-mass galaxies are expected to have a significant impact, as they ionize the gas surrounding them in the cosmic web, and enrich it with heavy elements produced by star-formation \citep{Bouwens2016, Dayal2018, Maiolino2019, Bacon2021, Atek2022}.

This enriched gas comprises the intergalactic medium (IGM) and circumgalactic medium (CGM), which consists of diffuse, multiphase gas that lies within the virial radii of galaxies \citep{Tumlinson2017}. This tenuous reservoir plays a crucial role in regulating a galaxy's supply of gas, as stellar-driven feedback deposits enriched material beyond the disk that can recycle through the galaxy on a variety of timescales \citep{Fielding2017, Peroux2020, Chadayammuri2022, Pandya2022, Peroux2022}. However, studying the connections between low-mass galaxies and the CGM at high redshift is challenging because the gas structures are intrinsically faint. In addition, the multiphase nature of the gas can only be fully characterized by combining the results of sensitive emission and absorption line studies \citep[e.g.][]{Steidel2010, Crighton2011, Turner2014, Bielby2019, Lofthouse2020, Stott2020, Beckett2021, Cooper2021, Fossati2021, Zabl2021, Bacon2023}.

These observational challenges have made the study of quasar fields particularly insightful because the quasar's light is selectively absorbed by the IGM and CGM as it travels to the observer, which reveals the gas in and around galaxies that reside in field, cluster, and group environments \citep{Bechtold2001, Khare2013, Chen2020, Dutta2020, Dutta2021, Lofthouse2023}. In the past, these studies were limited by the requirement that sources needed to be preselected for targeting with long-slit or fixed aperture multi-object spectroscopy, which can bias the galaxy sample. This can be overcome with integral field units (IFUs) that provide a spectrum for every spatial pixel within the field of view, or by employing slitless spectroscopy. Slitless grism observations avoid the aperture effects associated with long-slit observations, and provide high signal-to-noise (S/N) spectroscopy of a field at the cost of spectral resolution. Specifically, the data are typically low dispersion ($\lambda$/$\delta\lambda$~$\approx$~100) in order to disperse the full spectral traces of objects onto the detector, and can require advanced data processing techniques to properly extract spectra for sources in crowded fields \citep{Pirzkal2004, Pirzkal2013, Pirzkal2017, Momcheva2016, Brammer2019}.

These types of observations have typically been conducted from the ground in the optical \citep{Smith1975, MacAlpine1977, Schneider1999, Salzer2000, Salzer2005} to avoid high nightsky background levels and contamination from telluric emission at longer wavelengths. However, investigations of optical emission lines at higher redshifts are possible from space using the near-infrared capabilities of the Hubble Space Telescope (e.g. \citealp{Atek2010, Atek2011, Brammer2012, Colbert2013, Schmidt2014, Momcheva2016, Estrada-Carpenter2019, Pharo2019, Pharo2020, Bowman2021, Simons2021, Noirot2022, Papovich2022, Backhaus2023}).

\subsection{The MUSE Ultra Deep Field (MUDF)}\label{ssec:mudf}

We have undertaken an ultra-deep, near-infrared (NIR), optical, near-ultraviolet (NUV), and X-ray spectroscopic and imaging survey of the quasar field P2139-443, with the goal of investigating galaxy evolution in different environments. This field hosts two bright ($m_r$~$\approx$~17.9 and 20.5) quasars at $z\approx$~3.22 that are separated by only $\sim$1$\arcmin$ on the sky, corresponding to a physical separation of $\sim$500 kpc at $z\approx$~3. These two quasars are identified as J214225.78-442018.3 (also known as Q2139$-$4434, and hereafter QSO-SE) at $z =$~3.221~$\pm$~0.004 and J214222.17-441929.8 (Q2139$-$4433, hereafter QSO-NW) at $z =$~3.229~$\pm$~0.003 \citep{Lusso2019}. With a third quasar at nearly the same redshift (Q2138-4427, $z \approx$~3.17), and a separation of $\sim$8$\arcmin$, this field is a rare quasar triplet system \citep{Francis1993, D'Odorico2002}. There is also a fourth quasar near QSO-SE, albeit at a lower redshift ($z\approx$~1.298), and together these quasars probe a rich galaxy group at $z\approx$~0.88, an IGM filament at $z\approx$~3.04, and a candidate protocluster at $z\approx$~3.22 \citep{Fossati2019}.

Our current observations of the MUDF involve a six-prong strategy to cover submillimeter, NIR, optical, NUV, and X-ray wavelengths in emission and absorption from $z \sim$~0~--~6. These consist of imaging and spectroscopy with: (1)~the Hubble Space Telescope (HST) Wide Field Camera 3 (WFC3; \citealp{Turner-Valle2004, Kimble2008, MacKenty2010}) and Wide Field and Planetary Camera 2 (WFPC2; \citealp{Holtzman1995a, Holtzman1995b}), (2)~the Very Large Telescope (VLT) Multi Unit Spectroscopic Explorer (MUSE; \citealp{Bacon2010}), (3)~ the High Acuity Wide field K-band Imager (HAWK-I; \citealp{Pirard2004, Casali2006, Arsenault2008, Kissler-Patig2008, Paufique2010, Siebenmorgen2011}), (4)~the VLT Ultraviolet and Visual Echelle Spectrograph (UVES; \citealp{Dekker2000}), (5)~the X-ray Multi-Mirror Mission (XMM-Newton; \citealp{Jansen2001, Struder2001}), and (6) the Atacama Large Millimeter/submillimeter Array (ALMA; \citealp{Wootten2009}). The relatively small separation of the bright quasars aids in collecting these multiwavelength data sets, as the region of interest ($\approx$~2\arcmin~$\times$~2\arcmin) fits almost entirely within a single field of view (FOV) for most of the instruments.

The survey was launched with our large European Southern Observatory (ESO) program (ESO PID 1100.A$-$0528, PI: M.~Fumagalli) using VLT MUSE to observe this region, which we have dubbed the MUSE Ultra Deep Field (MUDF). The observations were obtained between 15~August~2017 to 2~June~2022, and accumulated a total of $\sim$142 hours of on-sky data. These optical IFU observations are complemented by high-resolution spectroscopy of the two quasars using UVES on the VLT (ESO PIDs 65.O$-$0299, 68.A$-$0216, 69.A$-$0204, 102.A$-$0194), which provides absorption line spectroscopy of numerous structures between $z \approx$~0~--~3 that we will connect with the physical properties of galaxies observed in emission. The VLT MUSE and UVES observations are detailed further in \cite{Lusso2019} and \cite{Fossati2019}, and here we focus on the recently acquired and archival HST observations.

By combining emission and absorption line spectroscopy with high spatial resolution imaging, our program is designed to connect the gas and galaxies observed in emission with the gas seen through absorption within field, group, and cluster environments. This will provide a more complete census of their physical properties and reveal the role of low-mass galaxies in enriching the gas. Specifically, we aim to characterize the gas kinematics, density, temperature, dust content, and metallicity, and reveal how these physical properties are shaped by different environments at various redshifts.

In this third paper, we expand our study of the MUDF to incorporate recently acquired HST observations, and provide the custom-calibrated data and source catalogs for use by the community. We describe the program design and acquisition of the HST observations in \S\ref{sec:obs}, and detail our custom-calibration of the direct imaging in \S\ref{sec:driz}. In \S\ref{sec:phot} we describe our process for identifying sources in the images and measuring their photometric properties, and in \S\ref{sec:grism} we describe our simulation based methods for extracting the slitless grism spectroscopy. In \S\ref{sec:morph} we discuss our morphology measurements, and in \S\ref{sec:hlsps} we describe the High Level Science Products (HLSPs) that are publicly available through the Mikulski Archive for Space Telescopes (MAST) Portal. Finally, in \S\ref{sec:summary} we summarize the current status of the program and discuss our planned forthcoming studies. We adopt the AB magnitude system ($m_\mathrm{AB}$, \citealp{Oke1983}) throughout this study, and assume a standard cosmology with $\Omega_M$~$\approx$~0.3, $\Omega_\Lambda$~$\approx$~0.7, and $h$~$\approx$~0.7 \citep{PlanckCollaboration2020}.

\section{Observations}\label{sec:obs}

\subsection{Program Overview}\label{ssec:overview}

\begin{figure}[htb]
\centering
\includegraphics[width=\columnwidth]{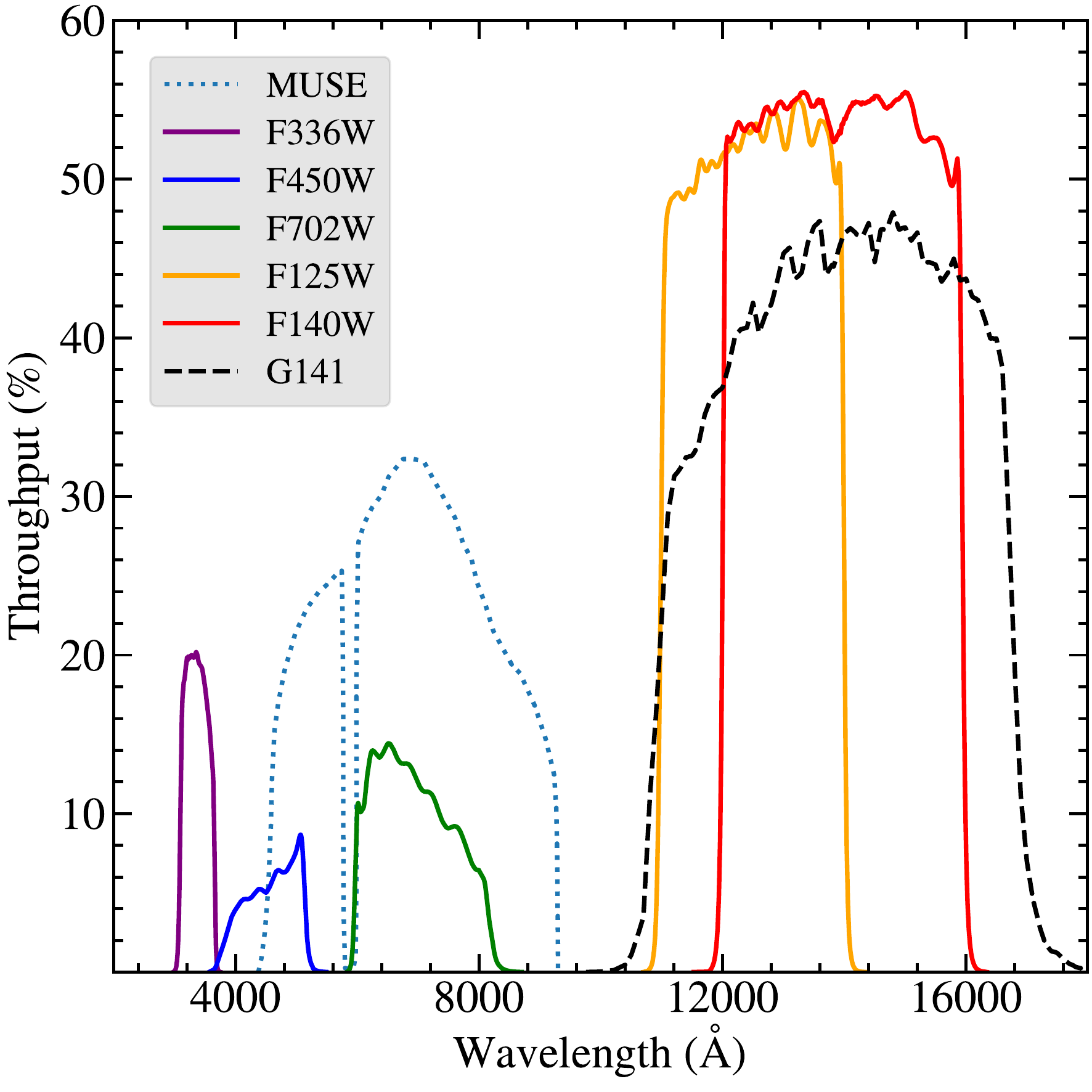}
\caption{The total system throughputs for the HST WFC3 (F140W, F125W, F336W) and WFPC2 (F702W, F450W) direct imaging filters, as well as the WFC3 G141 grism, and VLT MUSE. The transmission curves include the quantum efficiency of the CCD detectors.}
\label{fig:filters}
\end{figure}

\begin{figure*}[htb!]
\vspace{-1em}
\centering 
\includegraphics[width=\textwidth, trim={5em 5em 5em 5em}, clip]{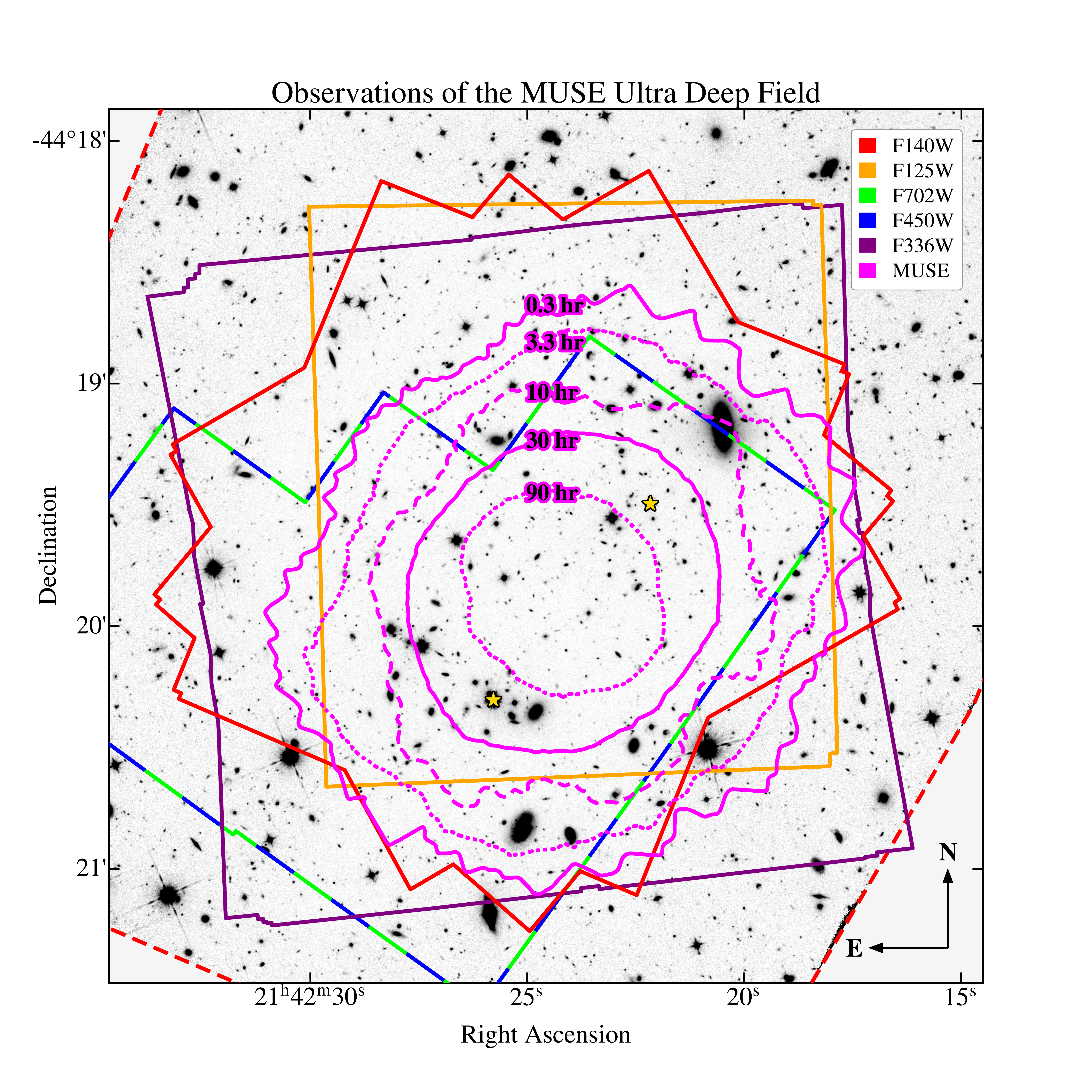}
\vspace{-1em}
\caption{The region of the sky (3$\farcm$6~$\times$~3$\farcm$6) containing the MUSE Ultra Deep Field, with instrument and filter coverage overlaid. The full field is covered by HST WFC3 F140W direct imaging, with a wide-shallow region (dashed red) and a narrow-deep region (solid red), with the data shown using a linear flux scaling. Additional HST WFC3 observations with the F125W (orange) and F336W (purple) filters are indicated by solid regions, as well as co-spatial archival WFPC2 observations with the F702W (dashed green) and F450W (dashed blue) filters. The depth of the MUSE observations are indicated by magenta contours that start at 90 hours and decrease by powers of three, with the center-most regions reaching a full depth of 120 hours. The locations of the two $z \approx$~3.22 quasars are indicated by gold stars, which are found within the 30~hour MUSE contour (solid magenta). Finally, the XMM-Newton and VLT HAWK-I observations cover areas larger than that shown in this graphic.}
\label{fig:obs}
\vspace{1.5em}
\end{figure*}

We first provide an overview of the HST observations in the MUDF, followed by a more detailed description of the program design and data acquisition. In the NIR, we have observed this field with WFC3/IR onboard HST, obtaining 90 orbits of G141 grism spectroscopy and F140W imaging (Program ID 15637, PI: M.~Rafelski \& M.~Fumagalli), resulting in the deepest HST grism survey ever conducted for a single field. The direct imaging consists of observations through the F140W filter to identify sources in the dispersed grism images, as well as with the F125W filter to constrain their spectral slopes. In the optical, we take advantage of archival HST WFPC2 imaging through the F702W and F450W filters (Program ID 6631, PI: P.~Francis), which provide useful constraints on the morphological and photometric properties of sources in the field. In the NUV, we have obtained eight orbits of HST WFC3/UVIS imaging using the F336W filter (Program ID 15968, PI: M.~Fossati). These observations probe the rest-frame UV emission of galaxies, allowing us to accurately characterize their star-formation histories. Together, the HST observations provide five-filter photometry (F140W, F125W, F702W, F450W, F336W) over the majority of the field, and the wavelengths sampled by each filter are shown in Figure~\ref{fig:filters}. Details of the observations are provided in Table~\ref{tab:obs}, and their on-sky locations are shown by outlined regions in Figure~\ref{fig:obs}.

\begin{deluxetable*}{lcccccccccccc}
\setlength{\tabcolsep}{0.02in} 
\tablecaption{Hubble Space Telescope Observations}
\tablehead{
\colhead{Instrument} & \colhead{Filter or} &\colhead{Effective} & \colhead{Zero} & \colhead{Number} & \colhead{Exposure} & \colhead{Image} & \colhead{Catalog} & \colhead{50\% Complete} & \colhead{Image} & \colhead{PSF} & \colhead{Align} & \colhead{Sources\vspace{-0.5em}}\\%
\colhead{and Camera} & \colhead{Grating} &\colhead{Center $\lambda$} & \colhead{Point} & \colhead{of Orbits} & \colhead{Time} & \colhead{Depth} & \colhead{Depth} & \colhead{Depth} & \colhead{Area} & \colhead{FWHM} & \colhead{RMS} & \colhead{Detected\vspace{-0.5em}}\\%
\colhead{} & \colhead{} &\colhead{(\AA)} & \colhead{($m_\mathrm{AB}$)} & \colhead{(N)} & \colhead{(sec)} & \colhead{(5$\sigma$ $m_\mathrm{AB}$)} & \colhead{(5$\sigma$ $m_\mathrm{AB}$)} & \colhead{($m_\mathrm{AB}$)} & \colhead{(arcmin$^2$)} & \colhead{($\arcsec$)} & \colhead{($\arcsec$)} & \colhead{(N)\vspace{-0.5em}}\\
\colhead{(1)} & \colhead{(2)} & \colhead{(3)} &\colhead{(4)} & \colhead{(5)} & \colhead{(6)} & \colhead{(7)} & \colhead{(8)} & \colhead{(9)} & \colhead{(10)} & \colhead{(11)} & \colhead{(12)} & \colhead{(13)}
}
\startdata
WFC3/IR & F140W (\textit{deep}) & 13,923 & 26.450 & 10 & 23,446 & 28.1 & 28.0 & 27.6 & 5.9 & 0.21 & 0.016 & 1,422 \\
WFC3/IR & F140W (\textit{wide}) & 13,923 & 26.450 & 1 & 2,100 & 27.5 & 27.2 & 27.0 & 17.9 & 0.21 & 0.021 & 1,953 \\
WFC3/IR & F125W & 12,486 & 26.232 & 1 & 2,120 & 27.1 & 26.5 & 26.9 & 5.0 & 0.21 & 0.013 & 1,248 \\
WFPC2 & F702W & 6,918 & 22.720 & 6 & 14,400 & 27.5 & 25.9 & 27.9 & 5.0 & 0.23 & 0.028 & 1,168 \\
WFPC2 & F450W & 4,557 & 21.921 & 8 & 19,400 & 27.0 & 25.4 & 27.6 & 5.0 & 0.19 & 0.027 & 1,092 \\
WFC3/UVIS & F336W & 3,355 & 24.687 & 8 & 20,890 & 29.1 & 29.9 & 28.4 & 7.5 & 0.11 & 0.004 & 1,580 \\
\hline
WFC3/IR & G141 & 13,875 & \nodata & 84 & 175,120 & \nodata & \nodata & \nodata & 5.9 & \nodata & \nodata & 1,499\\
\enddata
\tablecomments{A summary of the HST observations for this field, with all images drizzled to a pixel scale of 0$\farcs$06 pixel$^{-1}$ (60~milliarcseconds). The columns list: (1) the HST instrument and camera, (2) the filter (imaging) or grating (spectroscopy), (3) the filter pivot wavelength, (4) the AB magnitude zeropoint, (5) the effective number of orbits, (6) the exposure time in seconds, (7) the 5$\sigma$ isophotal image depth measured from empty sky regions using an aperture of 0$\farcs$2 in radius with uncertainties of $\pm$0.1 mag, (8) the 5$\sigma$ isophotal depth measured for compact objects in the catalog with S/N~=~4.5~--~5.5, (9) the 50\% completeness depth (see \S\ref{ssec:complete}), (10) the encompassed area on the sky, (11) the PSF FWHM, (12) the astrometric alignment RMS error relative to Gaia (for the F140W) and relative to the F140W mosaic (all other filters), and (13) the number of detected sources. The orbital visibility differs for each observation, so an effective number of orbits is quoted for the imaging by assuming an average exposure time of 2400~sec (40~min) per orbit. The F140W \textit{deep} region includes the \textit{deep} and \textit{wide} exposures, as they spatially overlap (see Figure~\ref{fig:obs}). The G141 grism observations cover a spectral range of 10,750~--~17,000~\AA~($\sim$1.1~--~1.7~$\mu m$) with a resolving power of \textit{R}~$\approx$~150 at 1.4~$\mu m$, where \textit{R}~=~$\lambda$/$\delta\lambda$ and $\delta\lambda$ is equal to twice the dispersion. Finally, non-applicable entries are indicated by (${}\cdots{}$). These data are available in MAST:\dataset[10.17909/q67p-ym16]{\doi{10.17909/q67p-ym16}}, together with our calibrated data products at:\dataset[10.17909/81fp-2g44]{\doi{10.17909/81fp-2g44}} and \url{https://archive.stsci.edu/hlsp/mudf}.}
\label{tab:obs}
\vspace{-2.5em}
\end{deluxetable*}

\subsection{Observational Setup}\label{ssec:obs}

The 90 orbits of WFC3/IR G141 grism spectroscopy and F140W imaging observations (Program ID 15637, PI: M.~Rafelski \& M. Fumagalli) were obtained in Cycle~26 between 10~August~2019 and 14~September 2019, with four failed visits repeated in March and September 2020. The program was designed with a standard image-grism-image exposure sequence for each orbit, which provides direct imaging in the F140W filter to identify the location of each object. This is required to measure the photometric and morphological properties of sources, and model how each object is dispersed across the detector during the grism observations so that exposures from different orbits can be aligned, and overlapping sources may be deblended during spectral extraction.

The observations were carried out at three fixed position angles (PAs) of \textsc{PA\_APER}~$=$~-150.3, -130.3, and -112.3 degrees, with the relative separations of $\sim$20 degrees to ensure that overlapping sources at one PA are well separated at other PAs. There are also sources outside the direct imaging field of view that are dispersed onto one side of the detector by the grism, so four orbits were allocated to wider-field imaging with the F140W filter using SPARS25 and NSAMP of 14. This wider imaging, combined with the primary images taken with each grism observation, provide the exact location of all sources with grism spectroscopy. Accordingly, the high-spatial resolution F140W imaging serves as the anchor for source detection and astrometric alignment of our other data sets in this field.

The observations were designed to maximize the depth of the grism spectroscopy in the central pointing by minimizing the background from zodiacal light and variable airglow from the He~I~$\lambda$10820~\AA~geocoronal line. This was accomplished by scheduling the observations with a minimum HST Sun Angle of 60 degrees, and excluding dates with low Earth Limb angles. In each orbit, a direct image was obtained, followed by six grism exposures that were split up in order to minimize persistence, followed by another direct image. The direct imaging employed SPARS10 with NSAMP of 10 to avoid saturating the quasars, resulting in exposure times of $\sim$93 seconds. The grism exposures used SPARS25 with NSAMP of 14, resulting in exposure times of $\sim$328 seconds. By bracketing the grism exposures with the direct images, we further reduced the effect of He~I airglow on the observations. Finally, the exposures were dithered using an extended six-point dither pattern to maximize the spectral resolution of the grism observations, and to improve the direct image quality by fully sampling the instrumental point spread function.

In addition to the HST WFC3/IR F140W imaging, we observed the field over a single-orbit with the F125W filter on 23~March~2020 to obtain photometry that better constrains the spectral slope of each object during extraction. This pointing is centered slightly to the north of the main F140W field, in order to fully capture objects that are dispersed onto the detector by the grism that lie outside the primary F140W field. We also pursued a complimentary program that obtained eight orbits of imaging with WFC3/UVIS using the F336W filter (Program ID 15968, PI: M.~Fossati). These observations were gathered during two visits on 5~May~2020 and 12~July~2020, and are critical for accurately modeling the star-formation histories of each galaxy. The F125W and F336W on-sky locations are shown by the orange and purple regions in Figure~\ref{fig:obs}.

Finally, we take advantage of existing WFPC2 optical imaging through the F450W and F702W filters (Program ID 6631, PI: P.~Francis, \citealp{Francis2001, Francis2013}) that were originally obtained to study galaxy clusters, and which provide us with useful constraints on the optical properties of many galaxies. The field was observed with the F450W filter for eight orbits between 31~August and 2~September~2000, while the F702W observations encompass six orbits on 23~August~1999. The F450W and F702W observations are co-spatial, as shown by the dashed blue and green regions in Figure~\ref{fig:obs}. There are also co-spatial observations through the F410M filter; however, this bandpass is fully encompassed by the much deeper F450W observations so we do not utilize the F410M data.

\section{Direct Imaging}\label{sec:driz}

\subsection{Drizzling the F140W Master Reference Mosaic}\label{ssec:f140w}

We aim to recover the highest possible spatial resolution in the HST images, which requires optimally aligning and combining the exposures to account for undersampling of the CCD detectors. Specifically, the telescope provides higher spatial resolution images than the detectors are able to record, which can be alleviated by making subpixel dithers between exposures. The maximum possible spatial resolution is then recovered by combining the images into a mosaic through a linear reconstruction process known as ``drizzling" \citep{Fruchter2002}. The Space Telescope Science Institute (STScI) provides a suite of tools for this within the \href{https://www.stsci.edu/scientific-community/software/drizzlepac.html}{\textsc{DrizzlePac}} software package \citep{Gonzaga2012, Hoffmann2021}.

First, a mosaic was produced by aligning and drizzling the HST WFC3/IR F140W images that were taken before and after each grism observation. The data were downloaded from the MAST archive on 11~March~2021, and were processed using the recently improved WFC3/IR Blob Flats \citep{Olszewski2021}, and the most recent bad pixel masks that were updated in the \href{https://hst-crds.stsci.edu}{CRDS} database the day prior. The 178 individual exposures were checked for quality, and 17 exposures with guiding or other issues were excluded. The remaining 161 exposures, each with an exposure time of $\sim$93 seconds, were then aligned and drizzled using the \textsc{TweakReg} (v1.4.7) and \textsc{AstroDrizzle} (v3.1.8) tools within \textsc{DrizzlePac} (v3.1.8).

In order to align the exposures, we first ran the \textsc{updatewcs} function with the \textsc{use\_db} option set to False. This removes the default WCS coordinates that cause the direct images and grism exposures to be misaligned, and restores the most recent HST guide star based WCS coordinates. The images were then aligned to the \href{https://www.cosmos.esa.int/web/gaia/early-data-release-3}{Gaia~EDR3} \citep{GaiaCollaboration2016, GaiaCollaboration2021} astrometric coordinate system using \textsc{TweakReg} by creating a reference catalog of Gaia sources with position uncertainties of $<$~10~milliarcseconds. We used the default \textsc{TweakReg} parameters, except we set the \textsc{imagefindcfg threshold} to 5$\sigma$ and the \textsc{peakmax} to 600 counts. There were typically 7~--~9 bright sources meeting these criteria in each exposure, allowing the images to be aligned to within an RMS error of only 16~milliarcseconds ($\sim$0.12~native pixels).
    
This yielded an intermediate drizzle consisting of the F140W exposures directly associated with the grism observations, indicated by the solid red region in Figure~\ref{fig:obs}. In addition, 28 longer exposures were obtained, split between four pointings (dashed red in Figure~\ref{fig:obs}), to capture objects that are dispersed onto the detector by the grism that lie outside of the primary F140W field. These each have an exposure time of $\sim$300 seconds, and were individually aligned to the intermediate drizzle using \textsc{TweakReg}. The default parameters were used, except that sources in each exposure were identified with \textsc{imagefindcfg} values of \textsc{conv\_width:3.5, threshold:4}, and \textsc{peakmax:1000}. The reference image parameters for \textsc{refimagefindcfg} were set to \textsc{conv\_width:5.0, threshold:6}, and \textsc{peakmax:250}. We employed the \textsc{use\_sharp\_round} option to identify the cleanest sources, which yielded $\sim$14~--~22 objects in each exposure. This process allowed us to align all of the F140W exposures to within an RMS error of 21~milliarcseconds ($\sim$0.16~native pixels).

These 189 exposures totaling 23,446 seconds ($\sim$6.5 hours) of on-sky time were then drizzled into a single master reference mosaic with a scale of 0\farcs06 pixel$^{-1}$. We optimized the \textsc{AstroDrizzle} parameters to produce a mosaic with the fewest artifacts and maximum depth. One criterion is that the standard deviation of empty background regions in the weight (WHT) map, divided by the median value in that region, should be $\leq$~0.2 (\S6.3.3 of the \textsc{DrizzlePac} Handbook v2.0; \citealp{Hoffmann2021}), which is controlled by varying the \textsc{final\_pixfrac} parameter. We also confirmed that the residual value in empty regions, which should be zero for a properly subtracted background, is $<$~0.0002 counts sec$^{-1}$. The adopted \textsc{AstroDrizzle} parameters are listed in Table~\ref{tab:driz}.

\begin{deluxetable}{ll}
\tabletypesize{\normalsize}
\tablecaption{{\normalsize \textsc{AstroDrizzle} Parameters}}
\setlength{\tabcolsep}{0.3in}
\tablehead{
\colhead{Parameter} & \colhead{Value}}
\startdata
\textsc{updatewcs} & False \\
\textsc{stepsize} & 1 \\
\textsc{skysub} & True \\
\textsc{skymethod} & globalmin+match \\
\textsc{combine\_type} & imedian \\
\textsc{final\_wht\_type} & IVM \\
\textsc{final\_pixfrac} & 0.6 \\
\textsc{final\_units} & cps \\
\textsc{final\_scale} & 0.06 \\
\textsc{final\_rot} & 0.0 \\
\enddata
\tablecomments{The \textsc{AstroDrizzle} parameters used to create the F140W mosaic (\S\ref{sec:driz}). These settings were also used for the other HST filters, except that the \textsc{final\_pixfrac} was set to 0.7 for the F702W and F450W filters, and the \textsc{combine\_type} was set to ``minmed" for the F702W filter.}
\label{tab:driz}
\vspace{-3.5em}
\end{deluxetable}

\subsection{The F125W, F702W, F450W, and F336W Mosaics}

We also produced drizzled images for the additional WFC3 and WFPC2 observations aligned to the F140W mosaic. The WFC3 F125W drizzle was produced using the same procedures as the F140W, except that the seven 303 second long exposures were aligned directly to the central F140W mosaic that is already astrometrically registered to Gaia. In the case of the WFPC2 F702W and F450W observations, we excluded short exposures with integration times of $\leq$~120 seconds, as these would degrade the S/N of the drizzle. The F702W drizzle was produced in a similar manner as the other filters, with the exception that the background levels of the four individual detector chips were normalized to the median value of each exposure. This was accomplished by convolving each exposure by a Gaussian kernel with a FWHM of three pixels, masking all sources detected with $\geq$~1$\sigma$ certainty, and calculating the sigma-clipped median background value for all chips. The exposures were then individually equalized to their median values by subtracting the mean background of each chip from the median. The overall correction is small, with a typical dispersion for the individual chip background levels of $\sim$0.4~--~0.8 counts. This correction was unimportant and not applied to the F450W filter, but for the F702W it yields a more uniform background, which is important for accurate photometry of a few objects that span multiple detector chips.

Finally, we applied several custom-calibrations to the WFC3/UVIS F336W exposures to utilize recent improvements. First, we used our publicly available codes\footnote{{\normalsize\url{https://github.com/lprichard/HST_FLC_corrections}}}\textsuperscript{,}\footnote{{\normalsize\url{https://github.com/bsunnquist/uvis-skydarks}}} that (1)~flag negative divots adjacent to readout cosmic rays (ROCRs) in the data quality (DQ) arrays of each exposure that are otherwise overcorrected in the default pipeline, and (2)~equalizes the background of the four amplifiers to remove background discontinuities, as discussed earlier for the WFPC2 F702W observations. These procedures are described in detail in \citet{Prichard2022} (see their \S2.2.2 and Appendix~A), based on techniques from \cite{Rafelski2015}.

Next, we corrected the F336W exposures for cosmic rays using our publicly available code\footnote{{\normalsize\url{https://github.com/mrevalski/hst_wfc3_lacosmic}}} that is based on the \href{https://www.astropy.org}{Astropy} implementation of the widely used \textsc{LACOSMIC} procedure \citep{vanDokkum2001, McCully2018, McCully2019}. First, the procedure flags any negative pixels that are $>$~5$\sigma$ below the median background and replaces them with a random negative value drawn from the 1$\sigma$ distribution. Without this correction, \textsc{LACOSMIC} improperly flags these $\sim$0.3\% of pixels as CRs and replaces them with large groups of negative pixels. Next, \textsc{LACOSMIC} is run, and the procedure allows the user to detect cosmic rays in each exposure, and either replace them with best-fitting values, or flag these as bad pixels in the Data Quality (DQ) arrays. We found that flagging the values in the DQ arrays so that they are ignored by \textsc{AstroDrizzle} produces the cleanest drizzle for eight exposures. In cases with fewer exposures, the replacement method may be required. Using a conservative 6$\sigma$ threshold, this process typically flags $\sim$5\% of the pixels as cosmic rays.

The exposures for each filter were then aligned to the F140W reference image using \textsc{TweakReg}, with alignment uncertainties $\sim$0.1~--~0.3 native pixels (see Table~\ref{tab:obs}). We used the same \textsc{AstroDrizzle} parameters listed in Table~\ref{tab:driz} for all of the filters, except the \textsc{final\_pixfrac} was set to 0.7 for the F450W and F702W filters, and the \textsc{combine\_type} was set to ``minmed" for the F702W filter. The exposures in each filter were drizzled to produce mosaics at scales of 0\farcs06 pixel$^{-1}$, with additional properties summarized in Table~\ref{tab:obs}.

\subsection{Mosaic Post-processing}\label{ssec:corr}

Finally, we post-processed the mosaics with our publicly available \href{https://github.com/mrevalski/create_neg_rms_images}{\textsc{create\_neg\_rms\_images}} code\footnote{{\normalsize\url{https://github.com/mrevalski/create_neg_rms_images}}} that generates a negative (NEG) image, and a Root-Mean-Square (RMS) error image from the weight map (WHT; in this case the inverse variance map) of each mosaic, where RMS~=~$1/\sqrt{\mathrm{WHT}}$. This post-processing also replaces NaNs with zeros and infinite values with $1\times10^{10}$, which lie outside of the data region, but would otherwise cause issues in the analysis. We also corrected the RMS maps for correlated pixel noise, which arises because the original pixels were shrunk to a smaller grid before drizzling, such that adjacent pixels share flux and noise information. If this effect is neglected, then the S/N will be overestimated when detecting and measuring the photometric and morphological properties of each object. The mathematical formalism is detailed in \S3.3.2 of the \textsc{DrizzlePac} Handbook (v2.0, \citealp{Hoffmann2021}), which provides a noise scaling factor ($R$) based on the drizzled and native pixel scales, and the \textsc{final\_pixfrac} used for each drizzle. Using our drizzle parameters, the noise scaling factors are: $R$~=~1.733 (F140W and F125W), $R$~=~1.629 (F702W and F450W, neglecting the small planetary chip), and $R$~=~0.455 (F336W). In the last case, the value is smaller than unity because the drizzled pixel scale is larger than the native WFC3/UVIS pixel scale. Last, we visually inspected each drizzle, and for the F702W, F450W, and F336W mosaics, we used their weight maps to generate region files and mask the extreme edges of the data that contained noisy pixels. The F140W mosaic with these refinements serves as our reference image for source detection, and the alignment of our other data sets in this field.

\subsection{Point Spread Function Modeling}\label{ssec:psf}

The spatial resolution of a telescope gradually decreases at longer wavelengths, which must be accounted for when measuring photometry in different filters. The spatial resolution for each filter is characterized by its Point Spread Function (PSF), which is a measure of how an unresolved source is spread onto the detector after being distorted by the optical elements of the telescope and instrument. Using a model of the PSF for each filter, the spatial resolution of the images can be degraded to match that of the longest wavelength image, which ensures that photometric measurements in different bandpasses sample the same intrinsic spatial area on the sky.

We used our publicly available \href{https://github.com/mrevalski/hst_wfc3_psf_modeling}{\textsc{hst\_wfc3\_psf\_modeling}} code\footnote{{\normalsize\url{https://github.com/mrevalski/hst_wfc3_psf_modeling}}} to construct PSF models for each mosaic. This code uses our source catalog (\S\ref{sec:phot}) to identify bright, isolated stars in each mosaic, and extracts a subimage of each star for stacking. The subimages are then normalized to their peak values and are interpolated to a finer grid for more precise alignment. The subimages are fit with Moffat profiles to determine the exact centroid location of each star, and then aligned to a common centroid. Finally, we calculate the mean and median of the aligned subimages, reinterpolate the results back to the mosaic pixel scale, and normalize the models to a total flux of unity. When a sufficiently large number of stars is available, adopting the median rather than the mean provides a higher S/N PSF model that robustly rejects contaminating sources.

\begin{figure*}[htb]
\vspace{0.5em}
\centering
\includegraphics[width=\textwidth, trim={1.2em 0em 0em 0em}, clip]{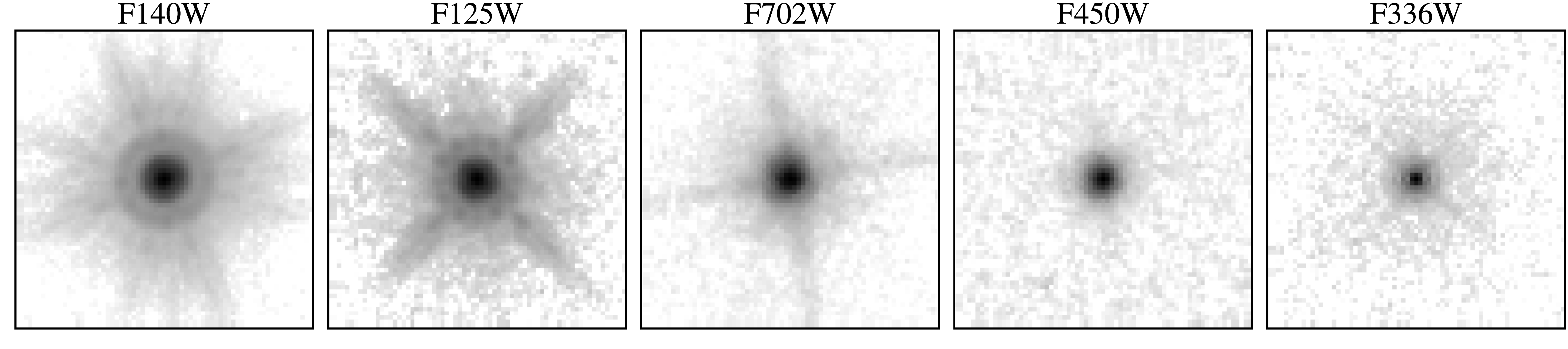}\\
\vspace{-0.8em}
\includegraphics[width=\textwidth, trim={1.2em 0em 0em 0em}, clip]{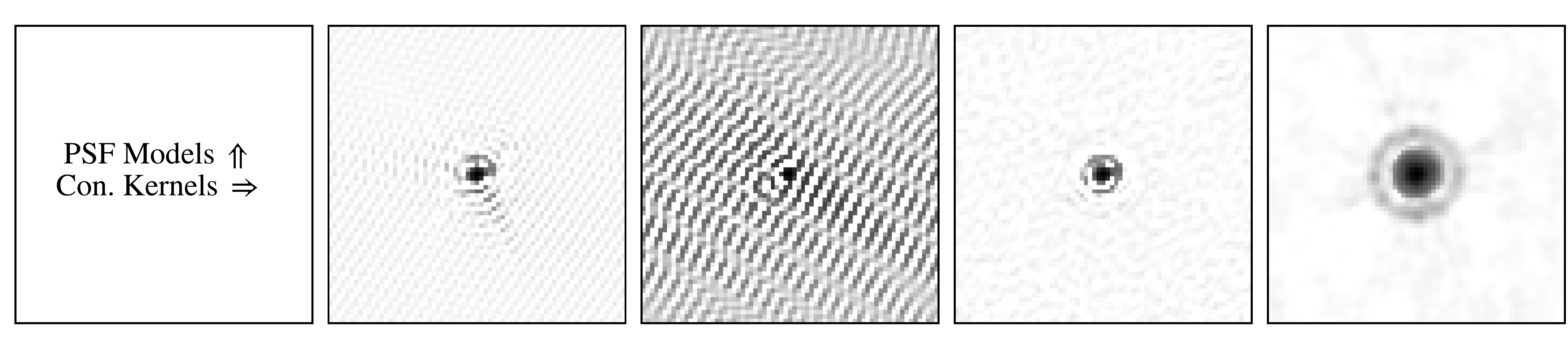}\\
\vspace{-1em}
\caption{The WFC3 (F140W, F125W, F336W) and WFPC2 (F702W, F450W) Point Spread Function (PSF) models (upper row) and convolution kernels (lower row) for the HST observations used in this study (see Table~\ref{tab:obs}). The PSF models are displayed on logarithmic scales, with their total fluxes normalized to unity. The convolution kernels were generated using a standard Hanning Window function, and are displayed on symmetrical logarithmic scales. As described in \S\ref{ssec:psf}, the convolution kernels are used to create lower spatial resolution versions of each mosaic that match the F140W to ensure that the photometric measurements in each filter use equivalent apertures that sample the same area on the sky.}
\vspace{1em}
\label{fig:psf}
\end{figure*}

In the case of the F140W and F125W images, nine stars\footnote{Catalog IDs: 1246, 1535, 1983, 2003, 2210, 2291, 20493, 20723, 21182.} were used in the median model stack, while six of these same stars were available for the F702W and F450W filters. Finally, four of these stars were used in the F336W stack. While a larger number of stars is desirable, the alternative is to use synthetic empirical PSF models, which do not accurately capture the extended wings of the PSF at the present time. The resulting PSF models are 4\farcs14~$\times$~4\farcs14 (69~$\times$~69 pixels) in size, and they are shown on logarithmic scales in Figure~\ref{fig:psf}.

We used these PSF models to generate convolution kernels that match (degrade) the spatial resolution of each drizzle to the F140W reference mosaic, which ensures that the photometric measurements in each filter are comparable. We created the kernels using the \href{https://photutils.readthedocs.io/en/stable/psf_matching.html}{PSF Matching} feature within Astropy's \href{https://photutils.readthedocs.io/en/stable/}{photutils} (v0.7) software package, and adopted a standard Hanning Window function, with the resulting convolution kernels shown in Figure~\ref{fig:psf}. We then convolved the images with their respective kernels using the \href{https://docs.scipy.org/doc/scipy/reference/generated/scipy.ndimage.convolve.html}{convolve} feature within the \href{https://www.scipy.org}{SciPy} (v1.3.1) multidimensional image processing (\href{https://docs.scipy.org/doc/scipy/reference/ndimage.html}{ndimage}) package. These PSF-matched mosaics are used for photometry with \textsc{Source Extractor} (\S\ref{sec:phot}), while the unaltered mosaics are used for morphologies with \textsc{statmorph}~(\S\ref{sec:morph}). In general, the differences for the photometry measured from the original and convolved images are negligible for bright sources, but become increasingly important for the faintest sources (e.g. for the F125W mosaic $\delta m \approx$~0.1~--~0.3 magnitudes).

\section{Source Catalog}\label{sec:phot}

\subsection{Source Detection}\label{ssec:detect}

The first step in analyzing the imaging, which is also required to extract the grism spectra, is to identify sources in the mosaics and measure their photometric and morphological properties. We selected \href{https://www.astromatic.net/software/sextractor/}{\textsc{Source Extractor}} (v2.5.0; \citealp{Bertin1996}) for this task, because it excels at detecting, segmenting, and deblending objects in crowded astronomical imaging. \textsc{Source Extractor} creates a segmentation map by detecting a minimum number of adjacent pixels with fluxes above a user-defined threshold, determining which groups of pixels belong to unique sources based on a deblending scheme, and then assigning a common integer identification number to the pixels in each group representing a unique object. We constructed a segmentation map for the F140W image by running \textsc{Source Extractor} with ``deep" and ``shallow" thresholds (also referred to as ``hot" and ``cold" in the literature) in order to better characterize the shapes and extents of faint and bright sources, respectively (see e.g. \citealp{Guo2013} and \citealp{Rafelski2015} for in-depth discussions).

The deep threshold is intended to detect the faintest sources, and was determined by running \textsc{Source Extractor} on the negative F140W image, which contains no true sources. The threshold is decreased incrementally, and the lowest value without false detections is adopted for the science image. The shallow threshold samples the brightest sources without encompassing nearby objects or extraneous background, and a value three times larger than the deep threshold was found to work well for this purpose. The final segmentation and photometry for each object is then selected from either the deep or shallow run based on a magnitude limit. In this way, bright sources are drawn from the shallow run, while faint sources are from the deep run. A limit of $m_\mathrm{F140W}$~=~19 was experimentally determined so that the segmentation of each object best matches the extent of its emission in the imaging. 

The deep and shallow segmentation maps and catalogs were then combined into single data products by taking objects brighter than $m_\mathrm{F140W}$~=~19 from the shallow run, replacing their entries in the deep segmentation map and catalog with the corresponding shallow values, and adding +20,000 to the ID number\footnote{The value of 20,000 is arbitrary, and any value larger than the number of sources in the deep run would suffice. In this case, there are $<$~10$^4$ sources, so a value of $>$~10$^4$ ensures an unambiguous identification for each source.} to indicate that these sources were drawn from the shallow run. The mosaic and merged segmentation map were then visually inspected to ensure that the segment for each object matched the physical extent of the source in the imaging, and five additional objects fainter than $m_\mathrm{F140W}$~=~19 were selected from the shallow run (IDs 588, 557, 780, 871, 1182) for this same type of replacement. Finally, we manually reassigned the IDs of diffraction spikes to match their sources for 10 bright stars drawn from the shallow catalog, and included their fluxes and error budgets in the photometry for these stars. Ultimately, this process selected 103 sources from the shallow run and 3,272 from the deep run, for a total of 3,375 sources. This procedure yields more reliable segmentation and photometry for very bright and faint sources than when using a single detection threshold, and the adopted \textsc{Source Extractor} parameters are summarized in Table~\ref{tab:se}.

\begin{deluxetable}{lcc}[hb]
\vspace{1em}
\tabletypesize{\small}
\tablecaption{\textsc{Source Extractor} Parameters}
\tablehead{
\colhead{Parameter} & \colhead{Deep Detection} & \colhead{Shallow Detection}
}
\startdata
\textsc{DETECT\_MINAREA} & 9 pixels & 6 pixels \\
\textsc{DETECT\_THRESH} & 0.6$\sigma$ & 1.8$\sigma$ \\
\textsc{ANALYSIS\_THRESH} & 0.6$\sigma$ & 1.8$\sigma$ \\
\textsc{DEBLEND\_NTHRESH} & 32 & 32 \\
\textsc{DEBLEND\_MINCONT} & 0.005 & 0.00001 \\
\textsc{FILTER\_NAME} & gauss\_3.0\_5x5 & gauss\_3.0\_5x5 \\
\textsc{WEIGHT\_TYPE} & MAP\_RMS & MAP\_RMS\\
\textsc{CLEAN\_PARAM} & 5 & 5 \\ \hline
Sources Extracted & 3,356 & 1,722 \\
\enddata
\tablecomments{The \textsc{Source Extractor} parameters used for all HST filters to create the MUDF catalog. The RMS maps were corrected for correlated pixel noise (see \S\ref{ssec:corr}), without which the equivalent deep and shallow thresholds are 1.1$\sigma$ and 3.0$\sigma$, respectively. We conservatively set the F140W and F125W \textsc{SEEING\_FWHM} parameter to 0\farcs16 for maximum detections and used the PSF FWHM values listed in Table~\ref{tab:obs} for the remaining filters. The images are background subtracted, and so no background parameters were specified. The \textsc{GAIN} parameter was set to the exposure time for each filter (Table~\ref{tab:obs}).}
\label{tab:se}
\vspace{-5em}
\end{deluxetable}

\subsection{Photometric Catalog}\label{ssec:cat}

The photometry for all filters was measured using \textsc{Source Extractor} in dual-image mode, where sources are detected in the deep F140W reference image, and then their fluxes are measured in the mosaic for each filter. If an object is detected above the user-specified thresholds (see Table~\ref{tab:se}), then \textsc{Source Extractor} measures the photometric parameters for that object. We used the same deep and shallow modes applied to the F140W, and the two catalogs for each filter were merged so that only the deep or shallow entry was adopted, matching the selection that was made for the F140W catalog.

\begin{figure*}[htb]
\vspace{0.5em}
\centering 
\includegraphics[width=\textwidth]{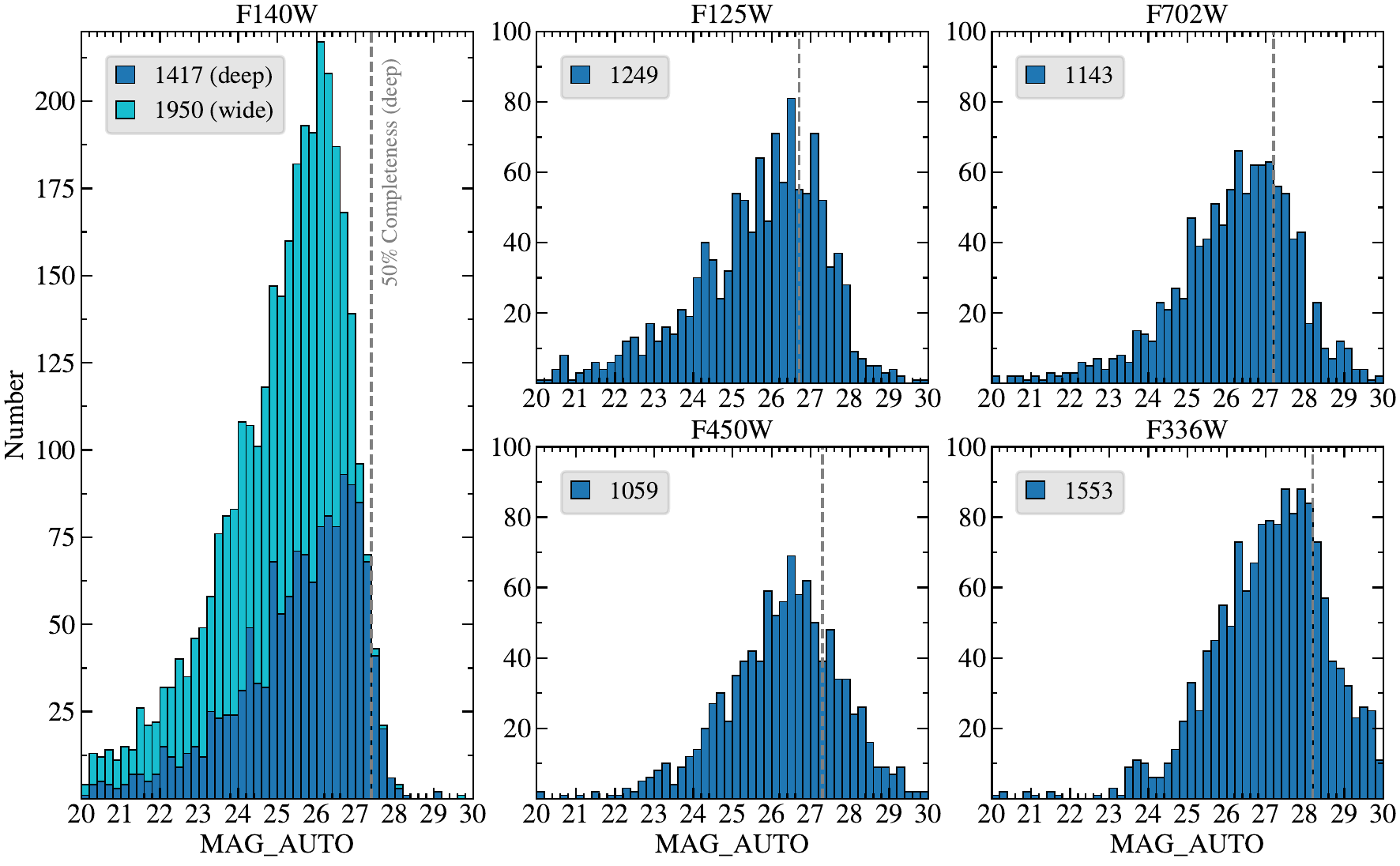}
\vspace{-1.5em}
\caption{Histograms of the automatic magnitudes (MAG\_AUTO) reported by \textsc{Source Extractor} for real sources in each HST mosaic. The number of sources is listed in each legend, and all panels use 0.2 magnitude bin widths. Sources in the deep and wide regions of the F140W mosaic (see Figure~\ref{fig:obs}) are stacked, and differentiated with blue and cyan, respectively. The brightest sources ($m_\mathrm{AB} <$~20) are typically foreground stars, while the faintest ($m_\mathrm{AB} >$~30) are noisy sources scattered above the detection threshold and so are not shown, and are considered upper limits. The vertical dashed lines represent the depths at which the detection efficiency reaches 50\% for unresolved sources, as shown in Figure~\ref{fig:complete}.}
\label{fig:mags}
\vspace{1em}
\end{figure*}

The measured fluxes in units of counts sec$^{-1}$ are converted to magnitudes by \textsc{Source Extractor} using the equation $m = m_\mathrm{ZP} - 2.5~\times$~log(flux), where $m_\mathrm{ZP}$ is the zeropoint for each filter (see Table~\ref{tab:obs}). The absolute flux calibration is determined by the calwf3 data processing pipeline, which produces the calibrated single-visit files (FLCs) that are combined during the drizzle process (\S\ref{sec:driz}). The HST instruments are exceptionally well characterized, and details of the calwf3 calibrations are described in Section 3 of the Instrument Handbook \citep{Sahu2021}, and \cite{Calamida2022}. The flux and magnitude values reported for each filter include the isophotal (\textsc{ISO}) and automatic (\textsc{AUTO}) aperture measurements. The isophotal values use the detection threshold as the lowest isophote, while the automatic measurements are based on the Kron radius \citep{Kron1980} that encompasses $\geq$~90\% of the total flux \citep{Infante1987, Bertin1996}, providing the most precise measurement for extended galaxies (Figure~\ref{fig:mags}).

The photometric measurements are PSF-matched across all filters, and the background level for each source is determined using a local aperture so that any residual spatial variations across the mosaics are handled self-consistently. We do not include any correction for Galactic extinction, which is very small for this field, with \textit{E}(\textit{B}--\textit{V}) $\leq$~0.018 (see \href{https://irsa.ipac.caltech.edu/applications/DataTag/}{NASA/IPAC IRSA Resolver} using the data tag: \verb|ADS/IRSA.Dust#2022/1110/082611_30964|).

The multiband photometric catalog was then assembled by merging the catalogs for the individual filters and selecting the columns of primary interest, with a total of 92 columns. The \textsc{Source Extractor} catalog is available in machine-readable form, and the catalog columns are listed in Table~\ref{tab:columns}.

\begin{deluxetable}{cl}[htb!]
\def\arraystretch{0.916}
\tabletypesize{\small}
\tablecaption{Source Catalog Columns}
\tablehead{
\colhead{Column Number} & \colhead{\textsc{Source Extractor} Parameter}
}
\startdata
(1) & NUMBER \\
(2) & X\_IMAGE \\
(3) & Y\_IMAGE \\
(4) & XMIN\_IMAGE \\
(5) & YMIN\_IMAGE \\
(6) & XMAX\_IMAGE \\
(7) & YMAX\_IMAGE \\
(8) & ALPHA\_J2000 \\
(9) & DELTA\_J2000 \\
(10) & A\_IMAGE \\
(11) & B\_IMAGE \\
(12) & THETA\_IMAGE \\
(13) & ELONGATION \\
(14) & ELLIPTICITY \\ \hline
(15) & KRON\_RADIUS\_F140W \\
(16) & FLUX\_RADIUS\_F140W \\
(17) & FWHM\_IMAGE\_F140W \\
(18) & CLASS\_STAR\_F140W \\
(19) & ISOAREA\_IMAGE\_F140W \\
(20) & MAG\_ISO\_F140W \\
(21) & MAGERR\_ISO\_F140W \\
(22) & FLUX\_ISO\_F140W \\
(23) & FLUXERR\_ISO\_F140W \\
(24) & MAG\_AUTO\_F140W \\
(25) & MAGERR\_AUTO\_F140W \\
(26) & FLUX\_AUTO\_F140W \\
(27) & FLUXERR\_AUTO\_F140W \\
(28) & BACKGROUND\_F140W \\
(29) & FLAGS\_F140W \\ \hline
(30 -- 44) & Same as cols. 15 -- 29 for the F125W filter \\
(45 -- 59) & Same as cols. 15 -- 29 for the F702W filter \\
(60 -- 74) & Same as cols. 15 -- 29 for the F450W filter \\
(75 -- 89) & Same as cols. 15 -- 29 for the F336W filter \\ \hline
(90) & SPEC\_Z \\
(91) & SPEC\_Z\_ERR \\
(92) & SPEC\_Z\_FLAG\\
\enddata
\tablecomments{The physical parameters included in the source catalog for the HST F140W, F125W, F702W, F450W, and F336W observations. The first seven columns provide the object ID number, the X and Y pixel locations, and their minimum and maximum extents. These are followed by the R.A. and Decl., major and minor axis sizes, orientation angle, elongation, and ellipticity. There are then 15 columns that repeat for each filter, containing the various radii, magnitude, and flux measurements defined in \textsc{Source Extractor} (\url{https://sextractor.readthedocs.io/en/latest/Measurements.html}). These include the isophotal (ISO) and aperture corrected (AUTO) values. The last three columns provide the derived redshift, redshift uncertainty, and redshift quality flag. The source catalog is available in machine-readable form (see \S\ref{sec:hlsps}).}
\label{tab:columns}
\end{deluxetable}

\subsection{Catalog Completeness}\label{ssec:complete}

We characterized the depths of the images and the completeness of our photometric catalog to empirically determine the fraction of objects that we may expect to detect as a function of apparent magnitude. First, we used the code described in \cite{Prichard2022}\footnote{{\normalsize\url{https://github.com/lprichard/hst_sky_rms}}} to determine the image depths. This procedure measures the RMS background value in 1000 randomly selected empty sky regions, generates a histogram of the RMS values, and then fits the distribution with a Gaussian profile to determine the sigma-clipped median for each image. The median RMS value is multiplied by the correlated pixel noise correction factor for that filter, and then converted to a limiting magnitude using the zeropoint and a specified aperture size. The correlated pixel noise corrections for each HST filter are listed in \S\ref{ssec:corr}, the zeropoints are provided in Table~\ref{tab:obs}, and in this case we utilized an aperture of 0$\farcs$2 in radius to calculate 5$\sigma$ limits. The results of this process are listed in Table~\ref{tab:obs}, and we note that the deep portion of the F140W mosaic reaches a 5$\sigma$ depth of 28.1~$\pm$~0.1 in AB magnitude, which is within $\sim$0.2 mag of the exposure time calculator estimate.

We also performed a standard completeness analysis by introducing artificial sources into our images at increasingly fainter magnitudes, and measuring their detection rates with \textsc{Source Extractor}. We used our PSF models for this purpose, which are already normalized to a total flux of unity, and so by convention have integrated magnitudes equal to the zeropoint of their respective filters. We then scale the PSFs to other magnitudes using a simple multiplicative factor. As in \cite{Fossati2019}, we then repeated these tests using mock galaxies convolved with the PSF, with S\'ersic indices of $n =$~2 inclined at 45$^{\circ}$ and effective radii of 0$\farcs$26. This size corresponds to $\sim$5~kpc at $z\approx$~1, which is typical of star-forming galaxies \citep{VanRossum2009}.

\begin{figure}[htb!]
\vspace{-0.25em}
\centering
\includegraphics[width=\columnwidth]{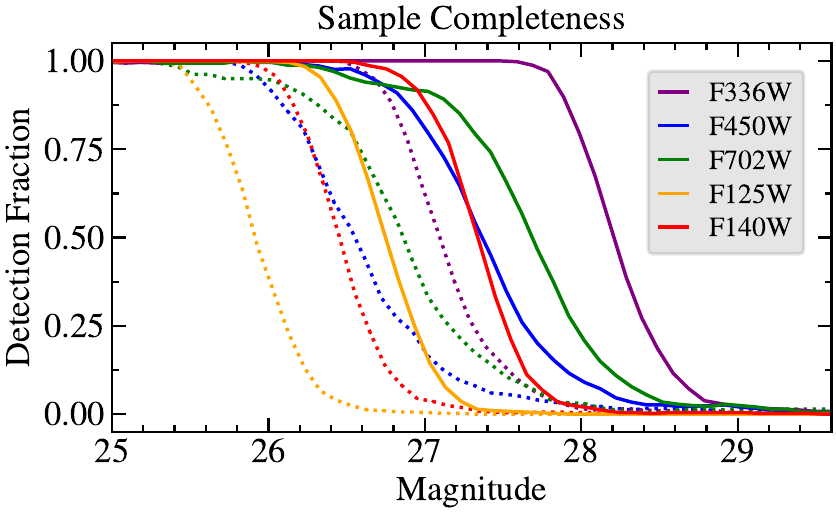}
\vspace{-1.75em}
\caption{The fraction of artificial sources that are recovered in each mosaic as a function of magnitude for point sources (solid lines) and extended galaxies (dotted lines). The sources were injected at random locations, with limits to avoid the extreme edges of the mosaics, the locations of real sources in the segmentation map, and the locations of other injected sources. The adopted limiting magnitude is the value at which the detection fraction decreases to 50\% (see Table~\ref{tab:obs}).}
\label{fig:complete}
\end{figure}

The results of this process are shown in Figure~\ref{fig:complete} and quantified in Table~\ref{tab:obs}, which suggest a slightly shallower limit than the RMS procedure. However, this is expected as the PSFs have FWHMs that are smaller than the 0$\farcs$4 diameter apertures used in the RMS measurements. The recovered magnitudes quoted in Table~\ref{tab:obs} are $\sim$0.15 mag fainter than the planted values shown in Figure~\ref{fig:complete}, which is consistent with measurements based on the Kron radius (see \S\ref{ssec:cat}). The F140W curve is shown for the deep region of the mosaic, and the wide portion follows a similar curve to that of the F125W, which is also one orbit in depth (see Table~\ref{tab:obs}). Despite comparable exposure times, the F336W is deeper than the F140W due to the lower background levels in the UV. However, galaxies are substantially brighter in the NIR, resulting in a higher S/N for detection.

\clearpage
\section{Grism Spectroscopy}\label{sec:grism}

\subsection{The Simulation Based Extraction Method}\label{ssec:sbe}

\begin{figure*}[htb!]
\centering
\includegraphics[width=\textwidth]{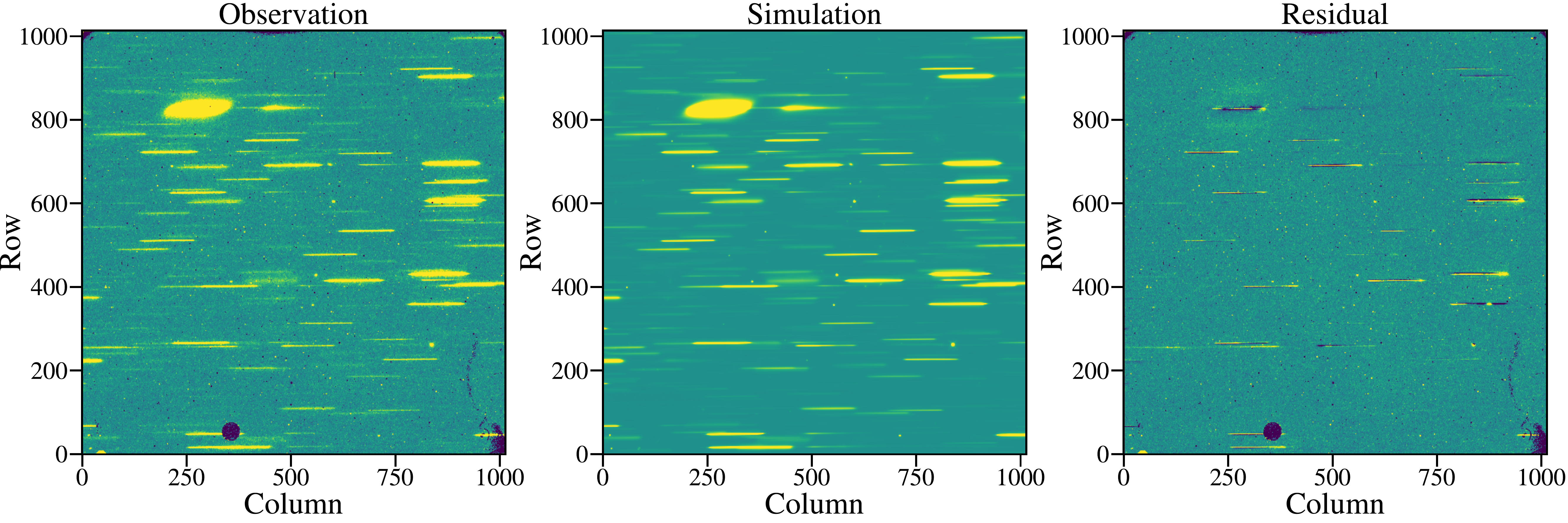}
\vspace{-1.75em}
\caption{The spectral modeling process for the HST WFC3 G141 grism data, using the example of a single exposure. The three panels show the data (left), simulation (center), and residuals (right). The small residuals for most objects highlight the accuracy of the astrometric alignment, the photometry in the simulations, and the subtraction of the dispersed background. The circular and curved regions of dead pixels near the bottom of the image are known as the ``death star" and ``wagon wheel" (see \citealp{Hilbert2009, Dahlen2010, Hilbert2010}).} 
\label{fig:grism}
\vspace{-0.5em}
\end{figure*}
\begin{figure*}[htb!]
\centering
\includegraphics[width=\textwidth]{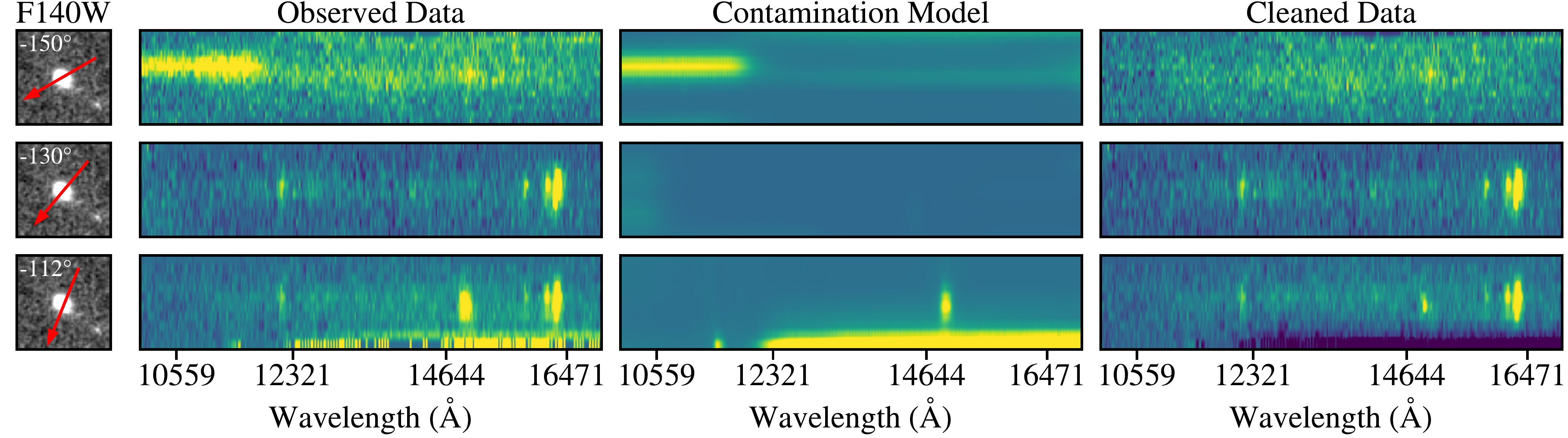}\\
\vspace{0.25em}
\includegraphics[width=\textwidth]{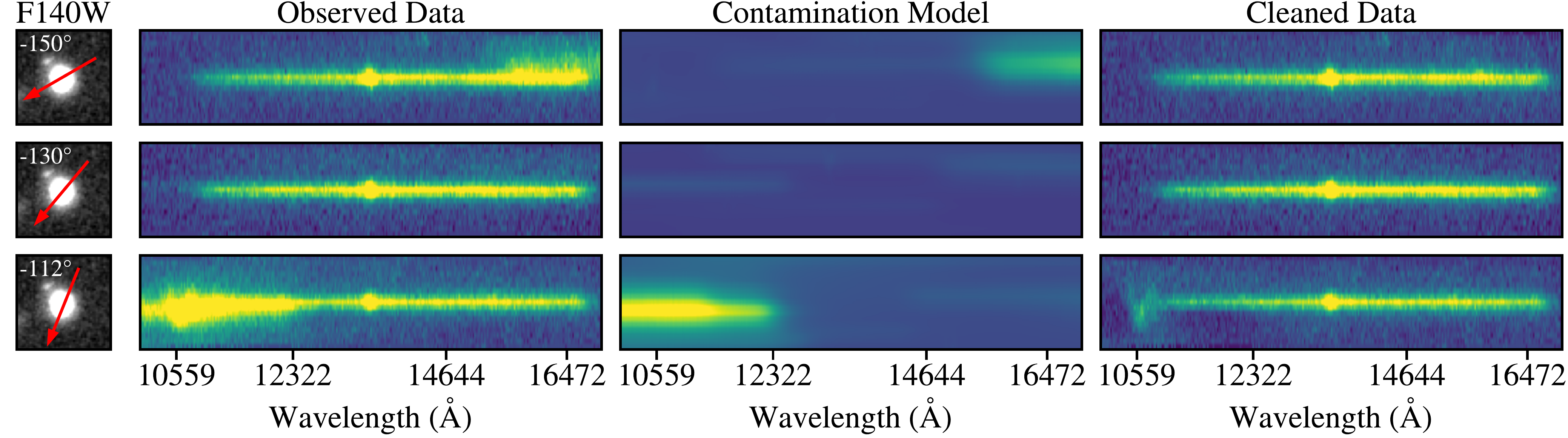}
\vspace{-1.75em}
\caption{The spectral extraction process for objects 1280 (upper rows) and 1388 (lower rows). The first column shows the F140W image and the dispersion angle for each grism spectrum (red arrows, -150$^{\circ}$, -130$^{\circ}$, and -112$^{\circ}$). The exposures taken at each orientation angle are stacked to produce high S/N 2D spectral images (second column), and the data are binned to 0$\farcs$129 in the spatial direction and 21.5~\AA~in the spectral direction to produce a spectrum with linear dispersion. The contamination model that accounts for emission from other sources (third column) is then subtracted from the data, resulting in a cleaned 2D spectral image (fourth column). The 1D spectra are then extracted using an optimal weighting scheme derived from the simulated images (see \citealp{Pirzkal2017}). The extracted 1D spectra for these objects are shown in Figure~\ref{fig:spectra}.}
\label{fig:grism2}
\vspace{0.25em}
\end{figure*}

\begin{figure*}[htb]
\vspace{0.5em}
\centering
\hspace{-0.01\textwidth}
\includegraphics[width=0.49\textwidth]{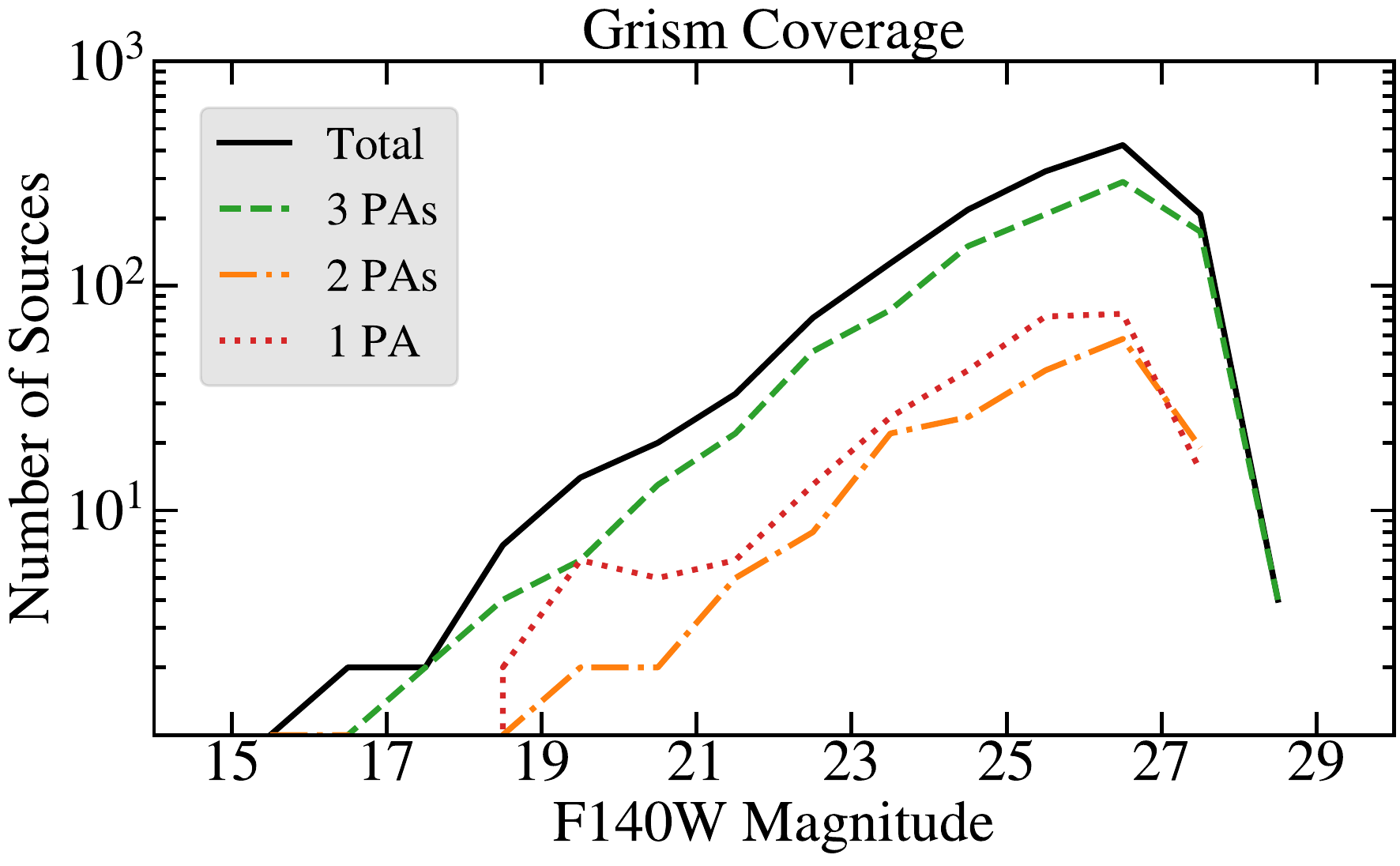}\hspace{0.01\textwidth}
\includegraphics[width=0.49\textwidth]{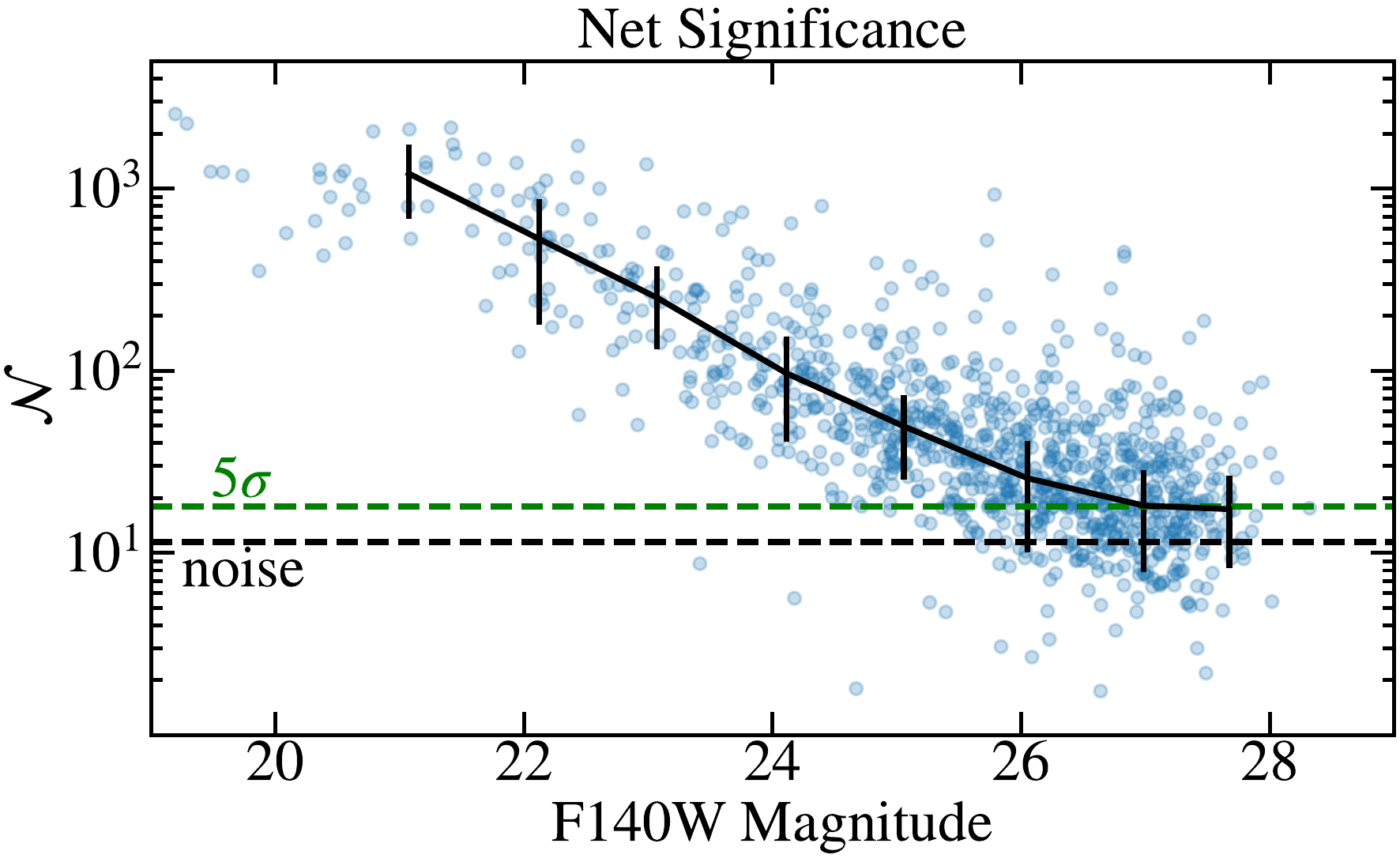}
\vspace{-0.5em}
\caption{The left graphic shows the number of extracted grism spectra with coverage at one (dotted red), two (dotted-dashed orange), or three (dashed green) position angles as a function of magnitude, with the total shown by the solid black line. The majority of sources ($\sim$70\%) have coverage in all three pointings. The right graphic shows the net significance ($\mathcal{N}$), or maximum cumulative S/N for each spectrum as defined in \S\ref{ssec:specdepth} for sources covered by three PAs. Overall, brighter sources have a higher $\mathcal{N}$ corresponding to higher S/N in the continuum and/or emission lines. The dashed black line corresponds to pure noise (no detection), while the dashed green line corresponds to a 5$\sigma$ detection (see \S\ref{ssec:specdepth} for details). In general, we detect the continuum emission of sources at $\geq$~5$\sigma$ confidence ($\mathcal{N}\approx$~18) down to a magnitude limit of $m_\mathrm{F140W}$~$\approx$~27.}
\label{fig:pas}
\vspace{0.5em}
\end{figure*}

We use the Simulation Based Extraction (SBE) method described in \cite{Pirzkal2017} to extract high-quality spectra that are flux and wavelength calibrated, free from contamination by other sources in any other spectral order, and reach the faintest possible emission line and continuum sensitivities. The SBE method relies principally on our ability to simulate each observation using a combination of photometric measurements and accurate calibration of the dispersion of the grism element. These simulations allow us to properly identify empty regions in each dispersed image so that the underlying background levels can be estimated accurately, and provide a quantitative estimate of the spectral contamination for each object that is removed during spectral extraction. These models also inherently contain the effect of the line spread function of the grism, which allows us to compute a sensitivity function specific to the morphology of each source. This applies to one-dimensional (1D) spectra that are integrated perpendicular to the dispersion axis, as well as two-dimensional (2D) spectra that contain both wavelength and spatial axes.

In general, the extraction of slitless spectroscopy requires knowing exactly where the spectral trace for every object is dispersed onto the detector, both to extract the flux of the object of interest, and to remove contamination of overlapping spectral traces from nearby sources. This requires an accurate measure for the position and size of every object in the field, which is provided by our source catalog in combination with accurate calibrations for the G141 grism dispersing element. The physical size of each object is important, because the shape of the dispersed spectrum is a convolution of the object's radial light profile with the intrinsic spectrum, i.e. emission lines for extended sources will be broader than for an unresolved object. While the grism observations were conducted at three PAs to model and remove contamination from overlapping spectral traces, this means that the 1D spectra at each PA for extended objects will differ in their shape and flux calibration. The SBE method accounts for this by using object-specific spectral responses derived from the data, simulations, and photometry, rather than assuming a point-source distribution (see \S4.2 of \citealp{Pirzkal2017}). Importantly, the F140W and F125W mosaics are used to derive a spectral slope for each object, which provides an important constraint on the relative fluxes of overlapping spectra during extraction.

We examined the 558 grism exposures and discarded 15 with poor image quality due to jitter during the orbit, or the presence of satellite trails that could not be fully masked. We used the latest available HST WFC3 G141 grism configuration files (see \citealp{Pirzkal2016}; WFC3-ISR 2016-15) and the improved background subtraction described in \cite{Pirzkal2020} (WFC3-ISR 2020-04). The latter allowed us to estimate and remove the variable He~I emission from the Earth's atmosphere. This He~I contamination varies during the course of the observations as described in \S\ref{ssec:obs}, and we carefully removed it before conducting on-the-ramp fitting of the multiple WFC3 reads of each exposure in order to produce a final calibrated image in units of electrons sec$^{-1}$.

\begin{figure*}[htb!]
\centering
\includegraphics[width=\textwidth, trim={5em 5em 5em 5em}, clip]{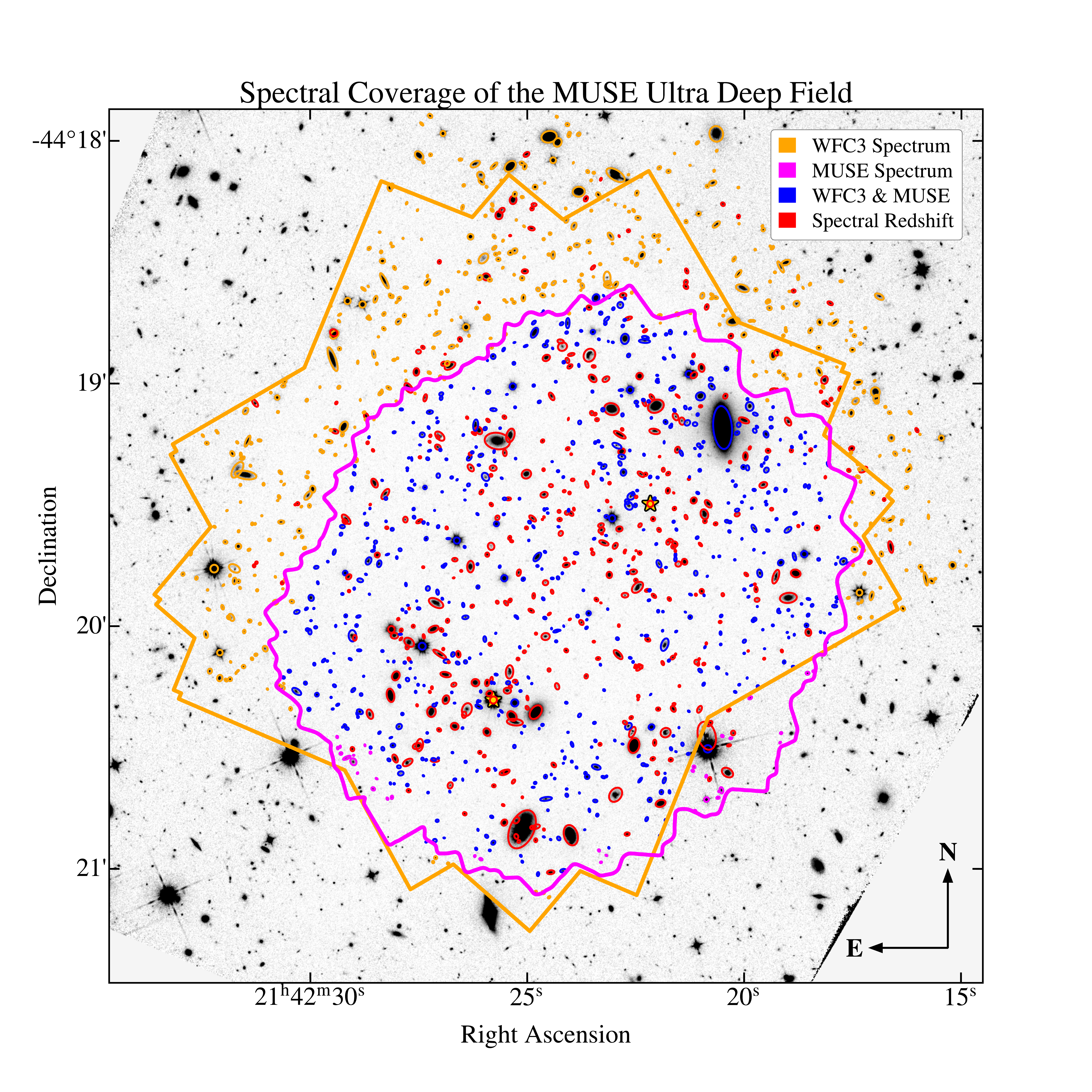}
\caption{The same field of view as Figure~\ref{fig:obs}, showing sources with spectroscopic coverage in WFC3 (1,499 sources) and MUSE (991 sources). The maximum spatial extent of the WFC3 and MUSE spectral regions is shown by the thick orange and magenta regions, respectively. Objects with spectra are shown with apertures equal to six times their Kron radii, and are colored in the following order: sources in orange have only a WFC3 spectrum (550 sources), magenta have only a MUSE spectrum (42 sources), blue have both WFC3 and MUSE spectra (949 sources), and red have a spectroscopic redshift (419 sources). The WFC3 grism disperses the field such that some sources with spectra are located just outside the direct imaging FOV, as seen by sources near the top of the figure, and bright objects without redshifts are primarily foreground stars in our Galaxy. These data provide spectroscopic coverage in at least one instrument for 1,541 sources, yielding 419 confirmed spectroscopic redshifts.}
\label{fig:obs2}
\end{figure*}

We then extracted the grism data using the techniques described above with our segmentation map and source catalog (\S\ref{sec:phot}), resulting in G141 spectra for 1,499 sources. An example of the spectral extraction process is shown in Figures~\ref{fig:grism} and \ref{fig:grism2}, and properties of the extracted spectra for the entire sample are summarized in Figure~\ref{fig:pas}. The extracted spectra are sampled at 21.5~\AA~pixel$^{-1}$, which is twice the native dispersion, to take advantage of the many dither positions. These spectra represent all of the sources contained in the deep central region of the F140W imaging, as well as off-detector sources that are dispersed onto one side of the detector, as shown in Figure~\ref{fig:obs2}. Finally, several examples of the extracted grism spectra are shown together with the direct imaging in Figure~\ref{fig:spectra}.

\subsection{Net Significance}\label{ssec:specdepth}

We characterized the quality and sensitivity of our grism spectra by calculating the net significance ($\mathcal{N}$), or maximum cumulative S/N, for each spectrum. As defined by \cite{Pirzkal2004, Pirzkal2017}, the net significance is calculated by (1) dividing every flux value by its associated error to produce an array of S/N values, (2) sorting the original flux and error arrays in decreasing order of the S/N estimates, and then (3) computing the cumulative S/N using a successively larger number of bins until the maximum value is reached.

This single value encompasses the information content of a spectrum as compared to pure noise, where a low value implies that there is minimal information above the noise. A high value corresponds to a strong detection of the continuum emission, an isolated emission line feature, or a combination of these two cases. As the pixels are sorted by S/N prior to summation, even a spectrum of pure noise will yield a positive $\mathcal{N}$ value. Based on \citet{Pirzkal2004}, a spectrum of pure Gaussian noise with $n_\mathrm{pix}$ independent pixels will have $\mathcal{N} = C \times (n_\mathrm{pix} / 100)^{1/2}$, where $C \approx 6.35 \pm 0.72$ and $n_\mathrm{pix} =$~325 for our extracted 1D grism spectra. This implies that a spectrum with $C > 8.5$ in the above relation corresponds to detecting real signal in the data at a 3$\sigma$ confidence level. However, it is important to emphasize that contamination, detector artifacts, noise properties, and small imperfections in the background subtraction can shift $\mathcal{N}$ to higher or lower values than these theoretical thresholds.

We show the net significance of our spectra as a function of the F140W magnitude in Figure~\ref{fig:pas}. In general, brighter sources have higher values of $\mathcal{N}$ because the continua are well detected, while fainter sources approach $\mathcal{N} \approx 15$, corresponding to an $\sim$3$\sigma$ detection. These characteristics are expected, as the observations were designed so that the depth of the direct imaging matches the faintest sources that can be extracted with confidence. While even deeper imaging would be useful to characterize the contamination of sources detected at the $<$~3$\sigma$ statistical level, this would have significantly reduced the depth of the grism spectroscopy. The distribution shown in Figure~\ref{fig:pas} indicates that, on average, a 5$\sigma$ continuum detection limit in the grism data corresponds to a source magnitude of $m_\mathrm{AB}\approx$~27 in the F140W mosaic.

\subsection{Spectroscopic Redshifts}\label{ssec:specz}

We provide barycentric spectroscopic redshifts for the entire sample in the current source catalog for completeness, and we will describe the spectral fitting process in greater detail in a subsequent publication (Revalski et al. \textit{in preparation}). In brief, we simultaneously fit the WFC3 and MUSE spectroscopy in order to identify emission line galaxies, determine their spectroscopic redshifts, and measure emission line fluxes. The line fluxes can be used to calculate line ratios that are sensitive to the physical conditions in the gas, including the temperature, density, ionization level, and dust content. 

We adopted the interactive spectral fitting routine described in \cite{Henry2021}, and modified it to fit a total of 34 emission lines from UV to NIR wavelengths. First, all of the spectra are automatically analyzed with a continuous wavelet transform (CWT), which robustly identifies sources with at least one emission line above a user-defined threshold. This process alleviates the requirement of visually inspecting every spectrum, which is an extremely time-consuming process for large surveys. We found that this process is highly complete, recovering $>$98\% of the sources that are identified visually and through other automated line-finding techniques.

Next, a cubic spline is simultaneously fit to the continuum of the WFC3 and MUSE spectra, and the code makes an initial guess of the object's redshift based on the strongest emission line. The user can then easily change the guess to other emission lines, adjust the points used in the continuum fit, mask regions of contamination, and then either reject the fit or save the results to an emission line catalog. In the case of objects with only a single emission line, we used the absence of other lines in either WFC3 or MUSE to determine the correct redshift. Specifically, when H$\alpha$~$\lambda$6563~\AA~is observed in the grism, then [O~III]~$\lambda$5007~\AA~is typically detected in MUSE, while at higher redshifts [O~III] is in the grism and the [O~II]~$\lambda$3727~\AA~doublet is resolved in MUSE. We will release the emission line catalog in a forthcoming publication, and the resulting redshifts for the sample are shown in Figure~\ref{fig:redz}.

The spectral dispersion of WFC3 and MUSE differ by a factor of $\sim$20, so it is important to consider differences in the redshifts derived from fitting each dataset. In the Figure~\ref{fig:redz} inset, we show the difference in velocity between the H$\alpha$ and [O~II] emission lines for sources where both can be detected. The redshifts were constrained to match within $\delta z\leq$~0.00334 ($\sim$1,000 km s$^{-1}$), except for a few sources. In 83\% of cases the difference in velocity from H$\alpha$ versus [O~II] is $<$ 500~km~s$^{-1}$, in 98\% they agree within 1,000~km~s$^{-1}$, and in 99\% of cases they agree within 2,000~km~s$^{-1}$. The catalog redshifts are based on the resolved [O~II]~$\lambda$$\lambda$3727, 3729~\AA~doublet when available, followed by H$\alpha$~$\lambda$6563~\AA.

Finally, we assign quality flags to the redshift measurements to indicate their confidence levels using the values listed in Table~\ref{tab:flag}. We note that resolved doublets such as [O~II] in MUSE and [O~III] in WFC3 are treated as two lines because they provide relatively unambiguous redshift measurements.

\begin{deluxetable}{clc}[ht!]
\vspace{-1em}
\tablecaption{Redshift Quality Flags}
\tablehead{
\colhead{Value} & \colhead{Criteria} & \colhead{Source Count}}
\startdata
4 & 2+ lines with S/N $\geq$ 5 & 366 \\
3 & 2+ lines with S/N $\geq$ 3 & 22 \\
2 & 1 line with S/N $\geq$ 5 & 31 \\
1 & spectral coverage, but no redshift & 1088 \\
99 & no spectral coverage for the source & 1868 \\
\enddata
\tablecomments{The redshift quality flags used in the source catalog.}
\label{tab:flag}
\vspace{-1em}
\end{deluxetable}

\begin{figure*}[htb!]
\centering
\includegraphics[width=\textwidth, trim={0em 4em 0em 0em}, clip]{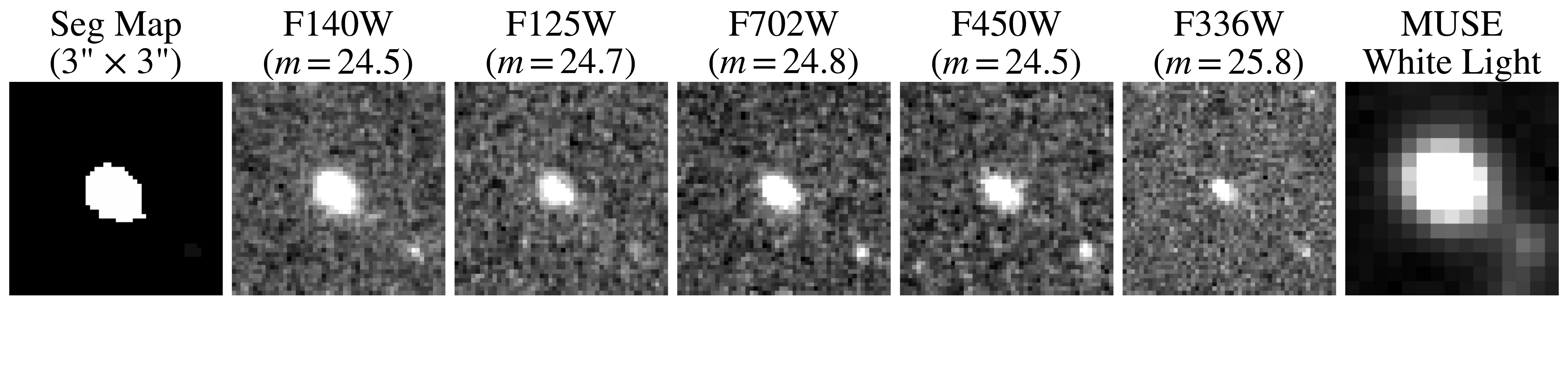}
\includegraphics[width=\textwidth]{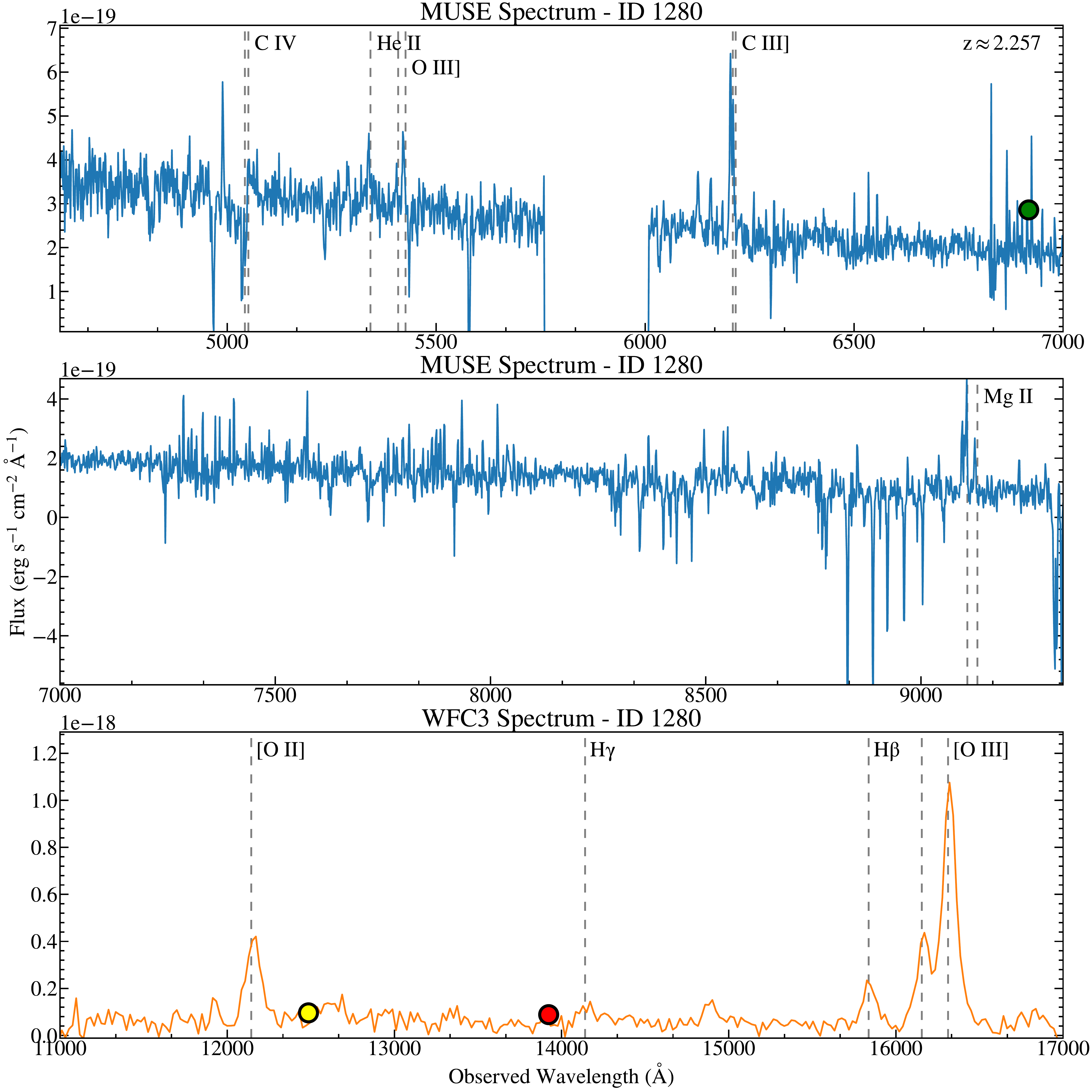}
\vspace{-1.5em}
\caption{The HST segmentation map, F140W, F125W, F702W, F450W, and F336W imaging, and MUSE white-light image (upper row), MUSE spectroscopy (middle rows in blue), and WFC3 G141 grism spectroscopy (lower row in orange) for object ID~1280, a luminous active galaxy based on emission line ratios, at a redshift of $z \approx$~2.257. The locations of detected emission lines are shown by vertical dashed lines, based on the best-fitting redshift value. The remaining calibration artifacts have negligible effects on the data quality, but include slight deviations in the WFC3 spectrum at the shortest wavelengths due to the decreased sensitivity of the grism, as well as residual skyline absorption features at the longest wavelengths in MUSE. The photometry for each filter is shown (when available) by a filled circle in the order: F450W (blue), F702W (green), F125W (yellow), and F140W (red). The fluxes are shown with 3$\sigma$ uncertainties, which are typically smaller than the size of the points.}
\label{fig:spectra}
\end{figure*}
\addtocounter{figure}{-1}
\begin{figure*}[htb!]
\centering
\includegraphics[width=\textwidth, trim={0em 4em 0em 0em}, clip]{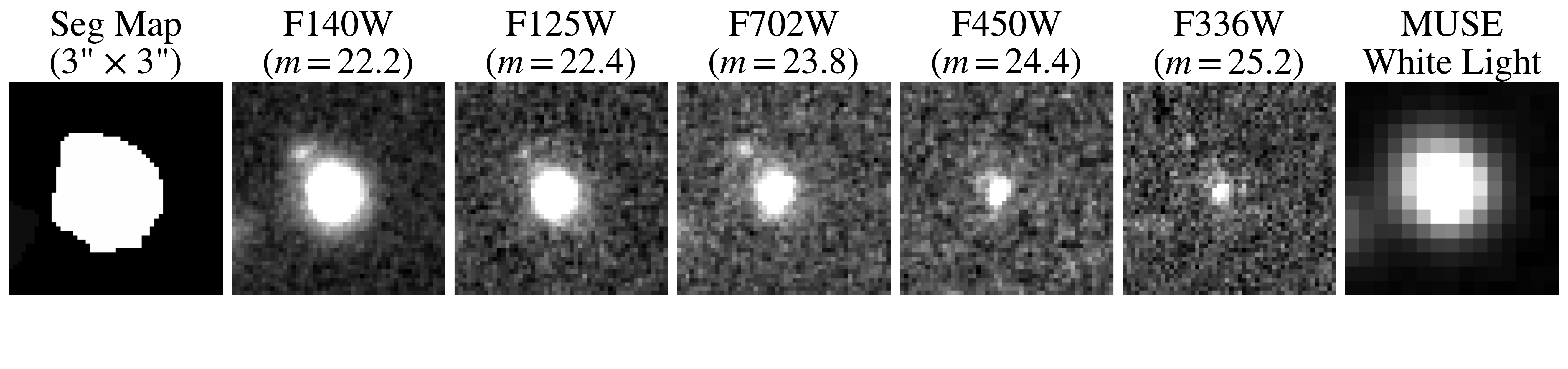}
\includegraphics[width=\textwidth]{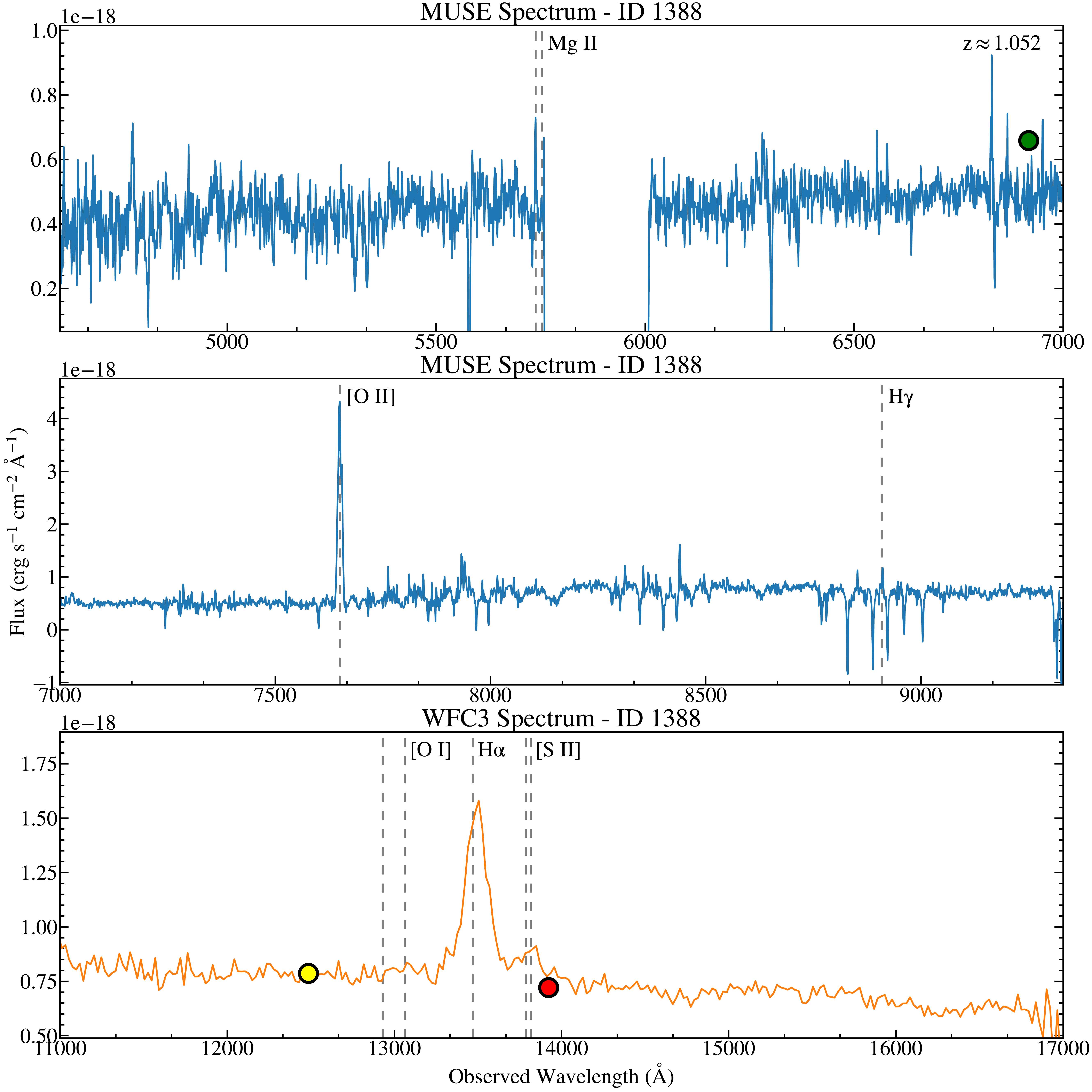}
\vspace{-1.5em}
\caption{{\it continued.}~The same as above for object ID~1388 at a redshift of $z \approx$~1.052.}
\end{figure*}

\begin{figure*}[htb!]
\vspace{1em}
\centering
\includegraphics[width=\textwidth]{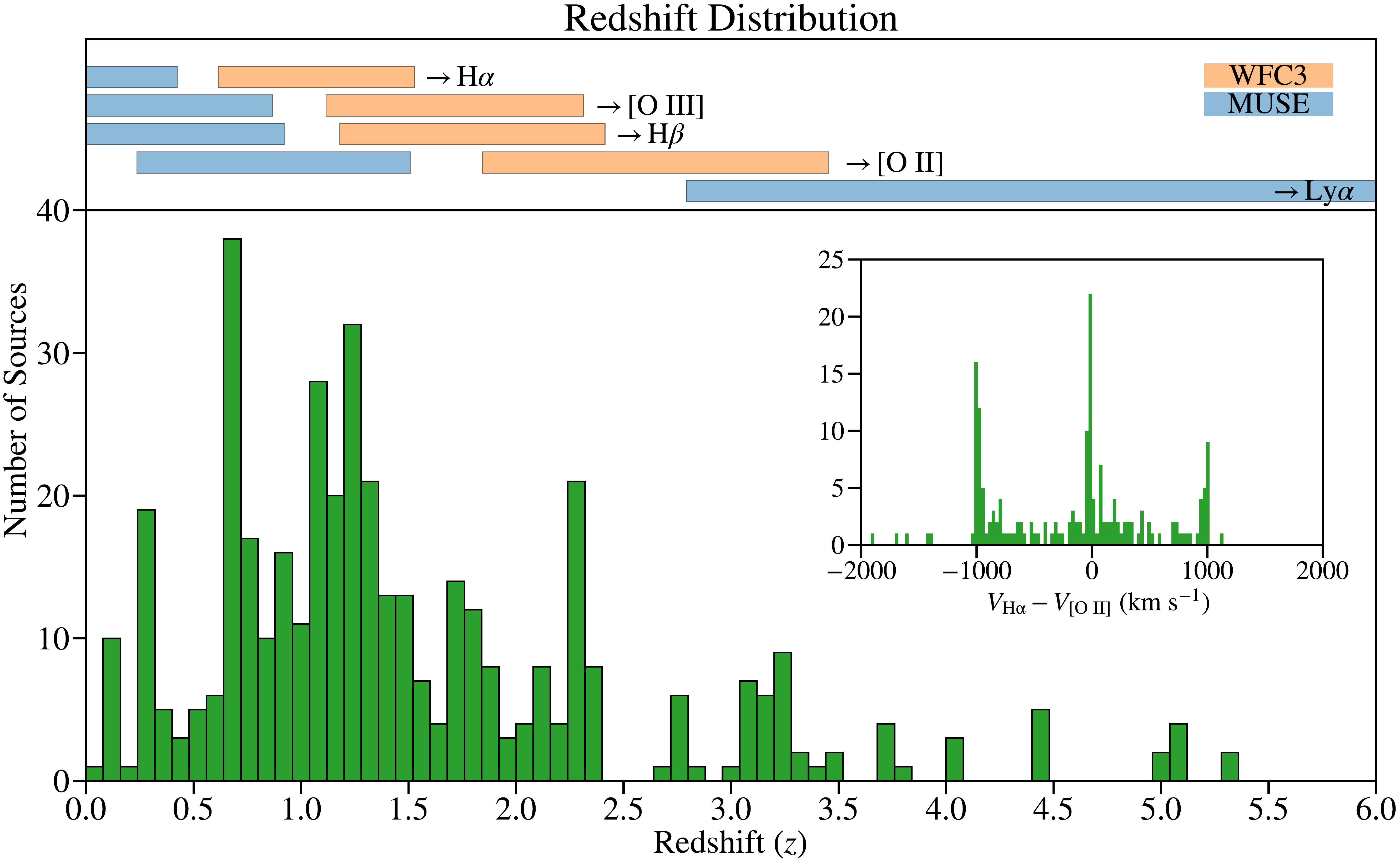}
\vspace{-1em}
\caption{The redshift distribution of sources in the MUDF catalog, with confirmed spectroscopic redshifts for 419 objects. The bins are $\delta z =$~0.08 in width, and sources between $z \approx$~0~--~3 are of particular interest because they potentially coincide with gas viewed in absorption from the two primary quasars at $z \approx$~3.22. The inset panel shows the differences in redshifts derived from the H$\alpha$~$\lambda$6563~\AA~line in WFC3 compared to the [O~II]~$\lambda$$\lambda$3727, 3729~\AA~doublet in MUSE for sources with detections of both emission lines (0.6~$<z<$~1.5). The redshifts of lines in MUSE and WFC3 were constrained to match within $\sim$1,000 km s$^{-1}$, except for a few sources where a larger offset improved the overall fitting result.}
\vspace{1em}
\label{fig:redz}
\end{figure*}

\textcolor{white}{.}
\newpage
\textcolor{white}{.}
\newpage
\textcolor{white}{.}
\newpage
\textcolor{white}{.}
\newpage

\section{Morphology Measurements}\label{sec:morph}

We used the Astropy \href{https://www.astropy.org/affiliated/index.html}{affiliated} package \href{https://statmorph.readthedocs.io/en/latest/}{\textsc{statmorph}}\footnote{{\normalsize\url{https://statmorph.readthedocs.io/en/latest/}}} (v0.4.0; \citealp{Rodriguez-Gomez2019}), to derive non-parametric morphological measurements for the sources in our photometric catalog. While \textsc{Source Extractor} measures some similar parameters, \textsc{statmorph} can also use a PSF model to measure more accurate sizes for compact sources, and provides additional diagnostics, such as the S\'ersic index \citep{Sersic1968}, as well as concentration, asymmetry, and smoothness (CAS) statistics \citep{Bershady2000, Conselice2003, Lotz2004}. The non-parametric formalism of \textsc{statmorph} has the advantage that many morphological parameters are reported without assuming a particular functional form for the underlying distributions of those parameters.

\begin{figure*}[htb!]
\centering
\includegraphics[width=\textwidth]{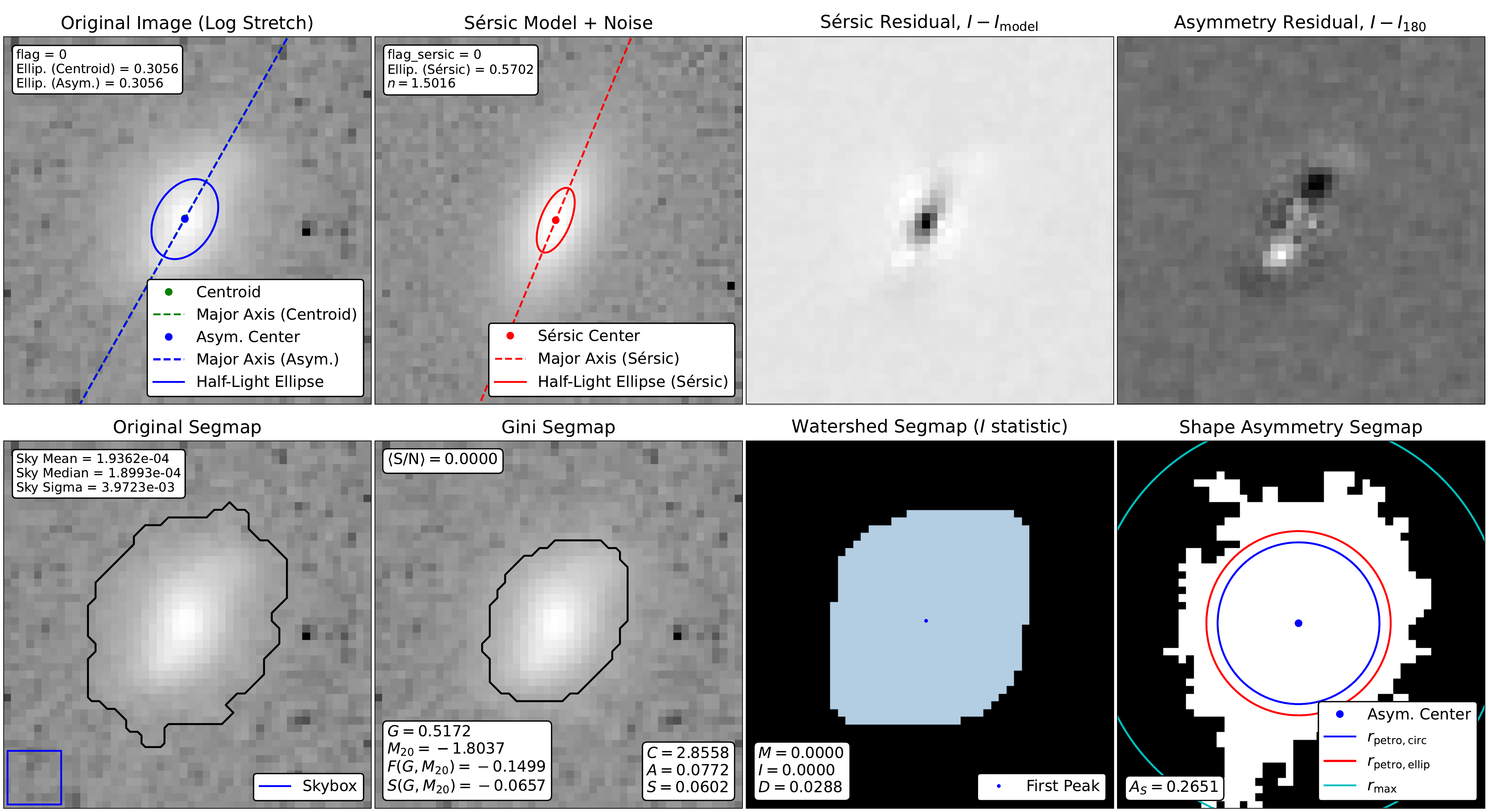}\\
\vspace{2em}
\includegraphics[width=\textwidth]{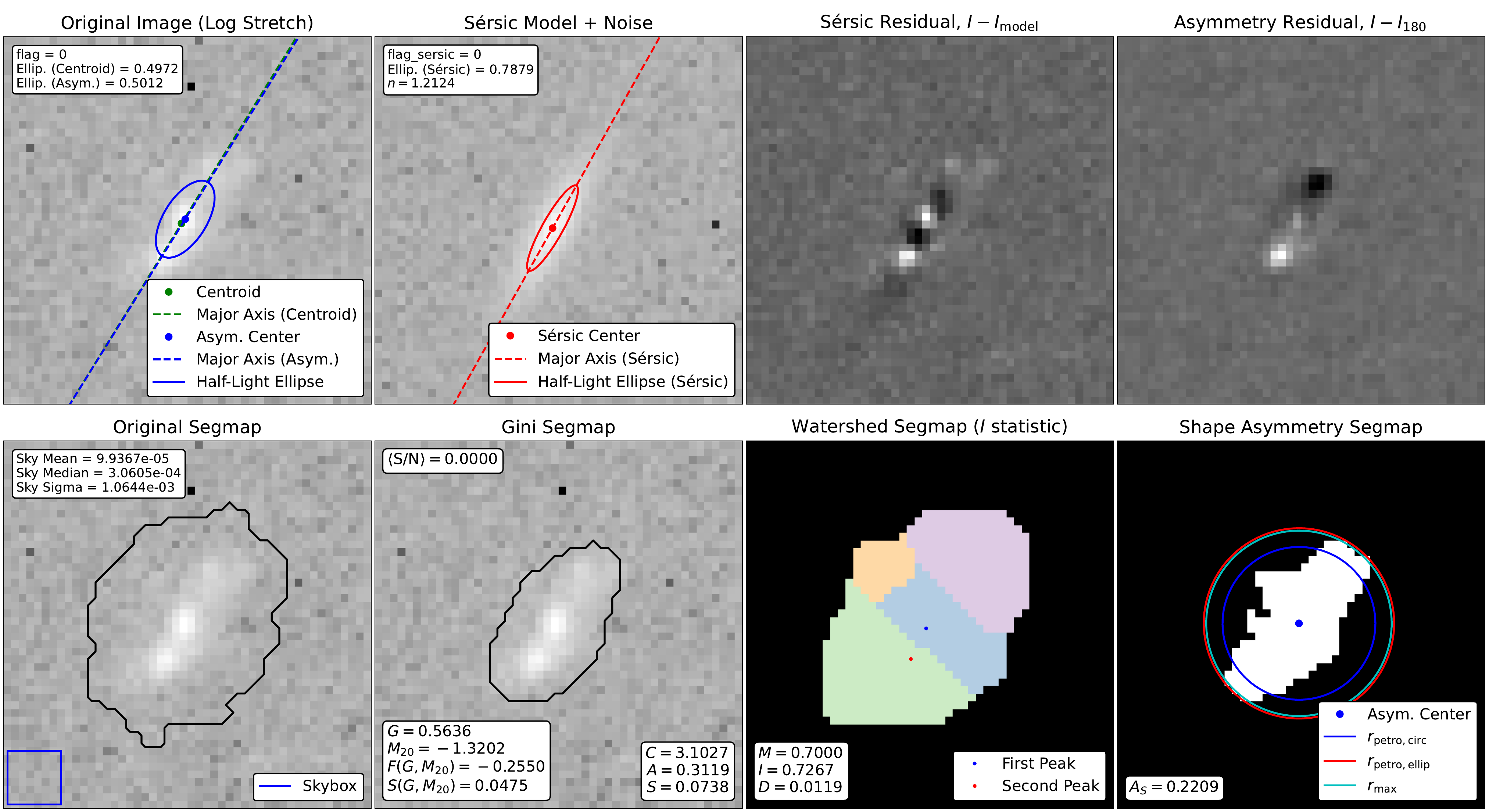}
\caption{An example of the \textsc{statmorph} results for object ID~2499 using the F140W (upper pane) and F336W (lower pane) filters. These figures are produced by the code's internal \textsc{make\_figures} function, and they summarize numerous morphological parameters for each object. The upper left quadrant contains two panels with the best-fit elliptical and S\'ersic models, and it displays the orientation and half-light radii for each model. These are subtracted from the image to generate residual maps (upper right quadrant). The lower left quadrant displays the original image with the user-supplied and Gini segmentation maps, while the lower right quadrant provides the watershed segmentation map where each color is associated with a local brightness maximum, followed by the shape asymmetry mask used to calculate the maximum and Petrosian radii. A complete description of the morphological parameters measured by \textsc{statmorph} is provided in Section~4 of \cite{Rodriguez-Gomez2019}.}
\label{fig:statmorph}
\end{figure*}

\begin{figure*}[htb!]
\centering
\includegraphics[width=\textwidth]{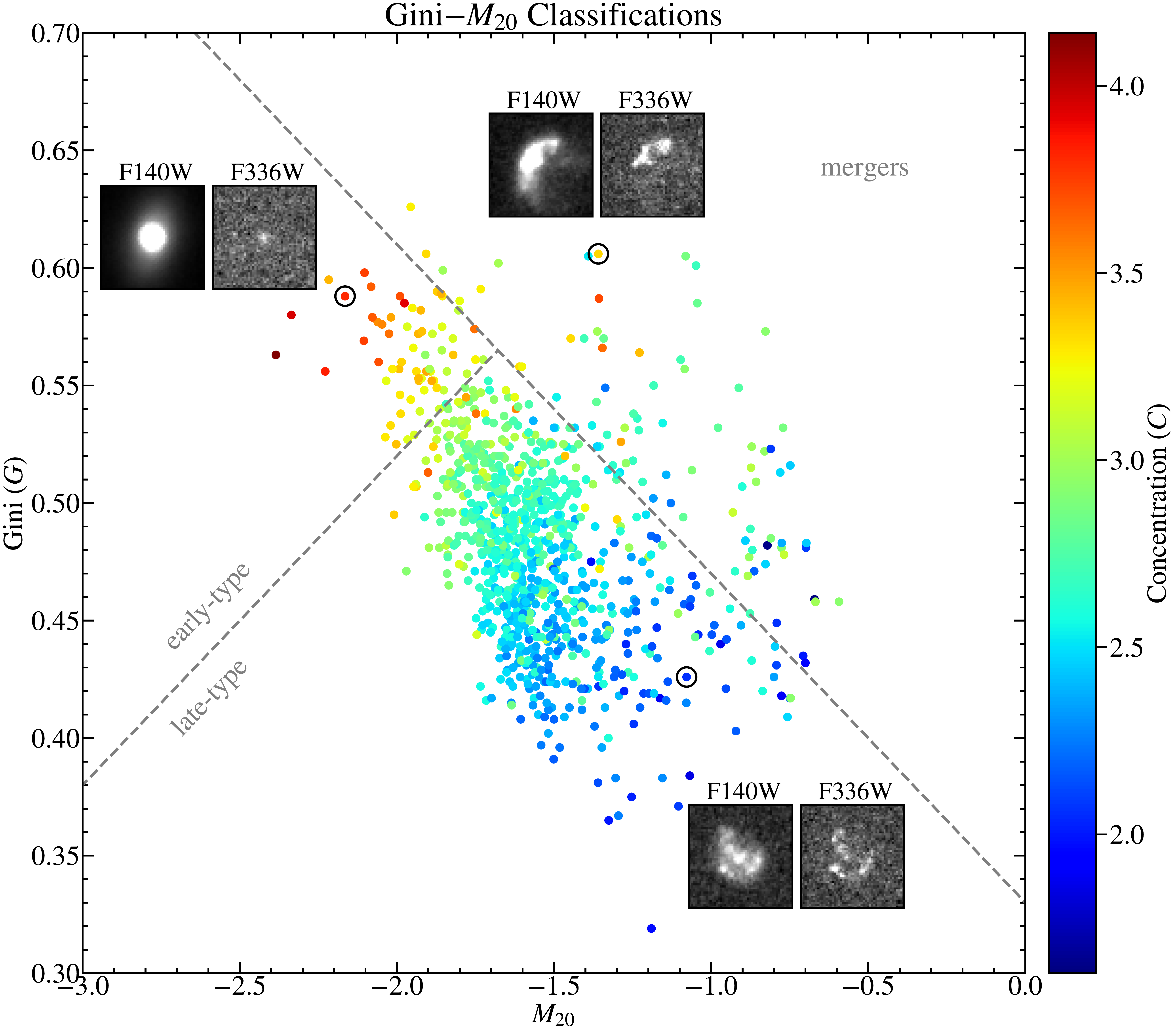}
\vspace{-1.5em}
\caption{The Gini$-M_{20}$ classification diagram for our sample, color-coded by the concentration index. The dividing lines for early-type, late-type, and merging systems are defined in \cite{Lotz2008} for systems at $z<$~1.2. Galaxies that are classified as early-type in the diagram are usually more concentrated based on their $M_{20}$ values. The inset panels show the F140W and F336W imaging for the sources in black circles.}
\label{fig:gm20}
\end{figure*}

An example of the morphological analysis for one object is shown in Figure~\ref{fig:statmorph}, with the significance of each parameter described in \cite{Rodriguez-Gomez2019}, and here we provide a brief overview. The code calculates a variety of size and shape measurements relative to the flux, asymmetry, and S\'ersic centroids of each object. The reported sizes include measures such as half-light radii ($r_\mathrm{half}$, $r_\mathrm{50}$), the Petrosian radius ($r_\mathrm{petro}$, \citealp{Petrosian1976}), and the second moment of each galaxy's brightest regions containing 20\% of the flux ($M_\mathrm{20}$). \textsc{statmorph} also provides concentration ($C$), asymmetry ($A$), and smoothness ($S$) statistics, along with the Gini index ($G$), which provide unique metrics for characterizing non-uniform distributions of light \citep{Conselice2014, Pawlik2014}.

We used the code with our segmentation map, full five-filter mosaics, and PSF models, in order to quantify how the morphologies of the sources change as a function of the observed wavelength. The F140W and F336W filters are particularly useful because they provide spatially-matched measurements of the rest-frame optical and UV emission for $\sim$1,580 sources across a range of redshifts. As described in \cite{Rodriguez-Gomez2019}, the code has several flags that report the quality of the fitted results for each object, with zero indicating success, and unity indicating a possible issue.

We provide the \textsc{statmorph} results in the form of machine-readable tables and code-generated figures that highlight key properties of the fit, including the object segmentation, major axis, S\'ersic index (when available), and CAS statistics. We only report the results for objects with a successful fit (\textit{flag}~=~0), with most of these also having a derived S\'ersic index (\textit{flag\_sersic}~=~0). Cases where the basic fit was successful, but a S\'ersic index could not be robustly derived as determined by the code, are indicated in the table with \textit{flag\_sersic}~=~1.

The parameters reported by \textsc{statmorph} can be used in combination to distinguish objects of different morphological types. One example is the widely used Gini-$M_{20}$ classification system \citep{Lotz2004, Lotz2008} that approximately separates early-type (E, S0, Sa), late-type (Sb, Sc, Sd, Irr), and merging galaxies, although how the Gini-$M_{20}$ plane is divided into these three regions varies across the literature (e.g. \citealp{Snyder2015, Rodriguez-Gomez2019}). We constructed this diagnostic diagram for our sample and show the results in Figure~\ref{fig:gm20}. As expected, galaxies that are classified as early-type based on their position in the Gini-$M_{20}$ diagram tend to contain most of their light in the central regions, lending confidence to the \textsc{statmorph}-derived properties. These measurements provide crucial morphological information for understanding the properties of galaxies in different environments, particularly as a function of observed wavelength, which will be further investigated in future work.

\begin{deluxetable}{cl}[htb!]
\def\arraystretch{0.86}
\tabletypesize{\normalsize}
\tablecaption{{\normalsize Morphology Catalog Columns}}
\tablehead{
\colhead{Column Number} & \colhead{\textsc{statmorph} Parameter}}
\startdata
(1) & id\\
(2) & flag\\
(3) & flag\_sersic\\
(4) & flag\_catastrophic\\
(5) & xc\_centroid\\
(6) & yc\_centroid\\
(7) & elongation\_centroid\\
(8) & orientation\_centroid\\
(9) & ellipticity\_centroid\\
(10) & concentration\\
(11) & asymmetry\\
(12) & smoothness\\
(13) & rhalf\_circ\\
(14) & rpetro\_circ\\
(15) & sersic\_n\\
(16) & sersic\_theta\\
(17) & sersic\_ellip\\
(18) & sersic\_amplitude\\
(19) & sersic\_rhalf\\
(20) & gini\\
(21) & m20
\enddata
\tablecomments{The \textsc{statmorph} parameters included in our catalog, with the exact definitions provided in \S4 of \cite{Rodriguez-Gomez2019}.}
\label{tab:statmorph}
\vspace{-3.25em}
\end{deluxetable}

\section{High Level Science Products}\label{sec:hlsps}

In Table~\ref{tab:hlsps} we document the files that are publicly available as High Level Science Products (HLSPs) from the Mikulski Archive for Space Telescopes (MAST) Portal. The data products encompass three categories: imaging, spectroscopy, and catalogs. The imaging includes our science drizzles, PSF models, convolution kernels, PSF-matched drizzles, and the segmentation map. The spectroscopy includes the calibrated 1D spectra and the 2D spectral images. Our catalogs include the multi-filter photometric catalog created with \textsc{Source Extractor} and the \textsc{statmorph} morphological catalog. Detailed descriptions of the data reduction and analysis techniques used to produce these data products are contained in Sections~\ref{sec:driz}~--~\ref{sec:morph}, and their key characteristics are provided in Tables~\ref{tab:obs}~--~\ref{tab:statmorph}. We reiterate that users may wish to apply their preferred correction for Galactic extinction to the photometry, although it is very small for this field (see \S\ref{ssec:cat}). In Table~\ref{tab:hlsps} we provide a list of the filename extensions for each type of data product, with files provided in the Flexible Image Transport System (FITS) format. The catalogs are provided in simple tabular text file formats. These high level science data products can be accessed from the MAST archive using the DOI:\dataset[10.17909/81fp-2g44]{\doi{10.17909/81fp-2g44}}. The files are also described on the MAST project webpage: \url{https://archive.stsci.edu/hlsp/mudf}.

\begin{deluxetable*}{cl}[htb]
\def\arraystretch{1.0}
\vspace{1em}
\setlength{\tabcolsep}{0.05in}
\tabletypesize{\normalsize}
\tablecaption{{\normalsize MAST HLSP Filename Extensions}}
\tablehead{
\colhead{Extension} & \colhead{Description}
}
\startdata
\textsc{$^\star$drz\_sci} & The science drizzles with pixel scales of 0.06 arcseconds per pixel and units of counts per second.\\
\textsc{$^\star$drz\_wht} & The drizzle weight maps produced using the inverse variance map (IVM) option in AstroDrizzle.\\
\textsc{$^\star$drz\_rms} & The drizzle error maps that are defined as 1/$\sqrt{\mathrm{IVM}}$ and have been corrected for correlated pixel noise.\\
\textsc{$^\star$drz\_neg} & The science drizzles multiplied by --1.0 so they may be used to determine source detection thresholds.\\
\textsc{$^\star$psf\_mod} & The area normalized PSF models that are 69~$\times$~69 pixels in size with scales of 0.06 arcseconds per pixel.\\
\textsc{$^\star$con\_ker} & The area normalized convolution kernels used to PSF-match each filter to the F140W mosaic resolution.\\
\textsc{$^\star$drz\_con} & The science drizzles convolved to the spatial resolution of the F140W mosaic and used for photometry.\\
\textsc{$^\star$seg\_map} & The merged deep + shallow segmentation map with IDs matching those in the merged source catalog.\\ \hline
\textsc{$^\star$1d\_spec} & The position angle combined HST WFC3/IR G141 1D spectrum for each object.\\
\textsc{$^\star$2d\_spec} & The position angle combined HST WFC3/IR G141 2D spectrum for each object.\\ \hline
\textsc{$^\star$source\_catalog} & The multi-filter \textsc{Source Extractor} object catalog with the columns listed in Table~\ref{tab:columns}.\\
{\footnotesize\textsc{$^\star$morphology\_catalog}} & The \textsc{statmorph} morphology catalog for each filter with the columns listed in Table~\ref{tab:statmorph}.\\
\enddata
\tablecomments{The files in the HLSP distribution (\url{https://archive.stsci.edu/hlsp/mudf}) consist of science images (\textsc{$^\star$.fits}), the multi-filter source catalog (\textsc{hlsp\_mudf\_source\_catalog}), and morphology catalogs (\textsc{$^\star$morphology\_catalog}). The science image filenames begin with \textsc{hlsp\_mudf\_hst\_wfc3\_$^\star$}, followed by the filter name, version number, and extensions listed above. The 1D spectral files contain multiple extensions (ext.) and we adopt \textsc{WAVELENGTH}, \textsc{WAVG\_OPT}, and \textsc{WSTD\_OPT} for our wavelengths, fluxes, and uncertainties, where \textsc{W$^\star$\_OPT} refers to the weighted, optimally-extracted values. The 2D spectral files contain the data, uncertainties, contamination model, and cleaned data in ext. 2~--~5, respectively, with \textsc{WAVELENGTH} in ext. 9.}
\label{tab:hlsps}
\vspace{-1em}
\end{deluxetable*}

\section{Summary}\label{sec:summary}

The MUSE Ultra Deep Field hosts two associated quasars at $z\approx$~3.22 that provide a unique laboratory for connecting the physical properties of galaxies seen in emission with their surrounding gas observed in absorption. We have observed this field over 90 orbits in a single pointing with the HST WFC3/IR G141 grism and F140W filter, comprising the deepest HST grism survey ever conducted for a single field. We combined WFC3 (F140W, F125W, F336W) and archival WFPC2 (F702W, F450W) imaging to detect 3,375 sources, with 1,536 having both spectroscopic and photometric coverage. The F140W and F336W mosaics reach depths of $m_\mathrm{AB}\approx$~28 and 29, respectively, providing near-infrared and rest-frame ultraviolet data for 1,580 sources, and we reach 5$\sigma$ continuum detections in the grism spectra for objects as faint as $m_\mathrm{AB}\approx$~27. By utilizing the long wavelength coverage of the MUSE and WFC3 spectroscopy, we derived robust spectroscopic redshifts for 419 galaxies between $z \approx$~0~--~6.

We have generated custom-calibrated science images, photometric and morphological catalogs, and high-legacy value data products for the MUSE Ultra Deep Field that are available to the community through the MAST Portal. These data products enable numerous studies aimed at advancing our models of galaxy formation and evolution in different environments. Specifically, we aim to connect the physical properties of the IGM and CGM traced in absorption with the galaxies discovered in emission to obtain a full census of the gas in and around galaxies in field, group, and candidate protocluster environments. Our forthcoming investigations include connecting the metal enrichment of the IGM and CGM with the gas phase metallicities of low-mass galaxies over cosmic time.

\acknowledgments

The authors would like to thank the anonymous reviewer for helpful comments that improved the clarity of this paper.

M. Revalski thanks Jay Anderson and Varun Bajaj for discussions on PSF modeling, Mihai Cara for assistance in drizzling WFPC2 data, Xin Wang for helpful discussions on PSF convolution window functions, Roberto J. Avila for Astropy WCS assistance, Edward Slavich for help in code parallelization, and Vicente Rodriguez-Gomez for \textsc{statmorph} support.

Based on observations with the NASA/ESA Hubble Space Telescope obtained from the MAST Data Archive at the Space Telescope Science Institute, which is operated by the Association of Universities for Research in Astronomy, Incorporated, under NASA contract NAS5-26555. Support for program numbers 15637 and 15968 was provided through a grant from the STScI under NASA contract NAS5-26555. These observations are associated with program numbers \href{https://archive.stsci.edu/proposal_search.php?mission=hst&id=6631}{6631}, \href{https://archive.stsci.edu/proposal_search.php?mission=hst&id=15637}{15637}, and \href{https://archive.stsci.edu/proposal_search.php?mission=hst&id=15968}{15968}. 

The MUSE portion of this project has received funding from the European Research Council (ERC) under the European Union's Horizon 2020 research and innovation programme (grant agreement No. 757535) and by Fondazione Cariplo (grant No. 2018-2329).

P. Dayal acknowledges support from the NWO grant 016.VIDI.189.162 (``ODIN") and the European Commission's and University of Groningen's CO-FUND Rosalind Franklin program and warmly thanks the Institute for Advanced Study (IAS) Princeton, where a part of this work was carried out, for their generous hospitality and support through the Bershadsky Fund.

This research has made use of NASA's Astrophysics Data System. This research has made use of the NASA/IPAC Infrared Science Archive, which is funded by the National Aeronautics and Space Administration and operated by the California Institute of Technology. This research made use of ccdproc, an Astropy package for image reduction \citep{matt_craig_2021_4588034}.

\facilities{HST (WFC3, WFPC2), VLT (MUSE), IRSA}

\newpage
\software{create\_neg\_rms\_images \citep{mitchell_revalski_2022_7458442}, hst\_wfc3\_lacosmic \citep{mitchell_revalski_2022_7458507}, hst\_wfc3\_psf\_modeling \citep{mitchell_revalski_2022_7458566}, Astropy \citep{AstropyCollaboration2013, AstropyCollaboration2018}, DrizzlePac \citep{Hoffmann2021}, Interactive Data Language (IDL, \url{https://www.harrisgeospatial.com/Software-Technology/IDL}), Jupyter \citep{Kluyver2016}, Matplotlib \citep{Hunter2007, Caswell2021}, NumPy \citep{Harris2020}, Python (\citealp{VanRossum2009}, \url{https://www.python.org}), Scipy \citep{Virtanen2020a, Virtanen2020b}, SExtractor \citep{Bertin1996}, Statmorph (\citealp{Rodriguez-Gomez2019}, \url{https://statmorph.readthedocs.io/en/latest/})}\\

Github repositories cited in this publication:
\begin{enumerate}
\vspace{-0.75em}
\setlength{\itemsep}{-0.25em}
    \item \url{https://github.com/mrevalski/create_neg_rms_images}\\(see: \cite{mitchell_revalski_2022_7458442},\dataset[10.5281/zenodo.7458442]{\doi{10.5281/zenodo.7458442}})
    \item \url{https://github.com/mrevalski/hst_wfc3_lacosmic}\\
    (see: \cite{mitchell_revalski_2022_7458507},\dataset[10.5281/zenodo.7458507]{\doi{10.5281/zenodo.7458507}})
    \item \url{https://github.com/mrevalski/hst_wfc3_psf_modeling}\\(see: \cite{mitchell_revalski_2022_7458566}, \dataset[10.5281/zenodo.7458566]{\doi{10.5281/zenodo.7458566}})
    \item \url{https://github.com/bsunnquist/uvis-skydarks}
    \item \url{https://github.com/lprichard/HST_FLC_corrections}
    \item \url{https://github.com/lprichard/hst_sky_rms}
\end{enumerate}

\newpage

\bibliography{references}{}
\bibliographystyle{aasjournal}
\end{document}